\documentclass[aps, prb, superscriptaddress, reprint]{revtex4-2}
\usepackage{amsmath}
\usepackage{color}
\usepackage{bm,graphicx,hyperref}
\hypersetup{%
  breaklinks = {true},
  citecolor = {blue},
  colorlinks = {true},
  linkcolor = {red},
}

\begin{document}

\title{Phonon Casimir effect in polyatomic systems}

\author{Gideon Lee}
\affiliation{Yale-NUS College, 16 College Avenue West, 138527, Singapore}

\author{Aleksandr Rodin}
\affiliation{Yale-NUS College, 16 College Avenue West, 138527, Singapore}
\affiliation{Centre for Advanced 2D Materials, 
			National University of Singapore, 
			6 Science Drive 2, 117546, Singapore}
	
\date{\today}

\begin{abstract}

The phonon Casimir effect describes the phonon-mediated interaction between defects in condensed-matter systems. Using the path-integral formalism, we derive a general method for calculating the Helmholtz free energy due to vibrational modes in systems of arbitrary dimensionality and composition. Our results make it possible to extract the defect interaction energy at any temperature for various defect configurations. We demonstrate our approach in action by performing numerical calculations for mono- and diatomic chains, as well as a diatomic molecule, at zero and finite temperatures and validate our results using exact diagonalization.

\end{abstract}	

\maketitle

\section{Introduction}
\label{sec:Introduction}

In his 1948 communication~\citep{Casimir1948ota}, Hendrik Casimir estimated the attractive force experienced by parallel conducting plates due to the electromagnetic vacuum fluctuations. He stated that ``although the effect is small, an experimental confirmation seems not unfeasible."~\citep{Casimir1948ota} The smallness of the effect was not exaggerated as it took fifty years of technological development to make the observation of this effect possible.~\citep{Lamoreaux1997, Bressi2002} In recent years, advances in nanotechnology~\citep{Krause2007, Klimchitskaya2009, Munday2009, French2010, Sushkov2011, Rodriguez2011, Zou2013, Intravaia2013, Garrett2019, Fong2019} and cold atoms~\citep{Moritz2003, Tolra2004,  Moritz2005, Catani2012} have stimulated the community's interest in this subject~\cite{Recati2005cfb, Bordag2009, Rahi2009, Reichert2018, Dehkharghani2018} as the relevant energy scales have become increasingly accessible in the experimental setting.

The phonon Casimir effect (PCE) is a related phenomenon describing phonon-mediated interaction between broken symmetry regions in solid-state materials. This symmetry breaking can be accomplished by replacing the lattice atoms with species of different masses, adding an external potential to restrict the motion of the system atoms, or modifying the force constant between some of the system's atoms. We will refer to all these modifications as ``defects." Instead of the electromagnetic vacuum, PCE is rooted in the zero-point phonon energy. Even though the expected energy scales are also expected to be small, recent estimates~\citep{Schecter2014pmc, Pavlov2019} suggest that cold atom techniques can be employed to investigate PCE.~\citep{Recati2005cfb, Schecter2014pmc}

What differentiates PCE from its electromagnetic counterpart is tunability. Because phonons are highly sensitive to the system's dimensionality and composition, it is, in principle, possible to engineer experimental setups to enhance the interaction energies. Therefore, a thorough understanding of PCE for general systems is desirable.

One of the earliest works on PCE~\cite{Schecter2014pmc}, focusing on pairs of dynamic impurities in one-dimensional systems with a single phonon branch, showed that at zero temperature and large impurity separation, the interaction energy between the defects decreases as the cube of the distance between them. Raising the temperature of the system results in an exponential suppression of the interaction if the separation exceeds the thermal de Broglie wavelength.~\cite{Schecter2014pmc}

Following this pioneering publication, the authors of Ref.~\cite{Pavlov2018pmc} performed a detailed study to demonstrate that the power-law dependence of the interaction is, in fact, a quasi-power-law with a variable exponent, approaching $-3$ at large separations. In addition to treating atomic impurities in 1D, the authors of Ref.~\cite{Pavlov2018pmc} addressed the interaction between lattice atoms confined by an external harmonic potential. The authors extended their analysis to two- and three-dimensional systems with a single phonon branch in Ref.~\citep{Pavlov2019} They showed that the interaction decays faster at higher dimensionalities and confirmed the temperature-induced interaction suppression at larger separations. Similarly to Ref.~\citep{Schecter2014pmc}, these works focused on two defects at a time.

In this work, we develop a general formalism that allows one to treat systems of any dimensionality with an arbitrary number of phonon branches and defects at any temperature. In fact, our method applies even to non-crystalline systems, where vibrational modes cannot be labelled by their crystal momentum. Using path integrals, we derive a formula for the Helmholtz free energy for systems hosting impurities, harmonic potential wells, and modified bonds, from which the interaction energy can be obtained. Unlike earlier work, where potential wells and impurities were treated separately, our approach makes it possible to mix various defect types. The ability to treat multiple defects simultaneously is important because the interaction between defects in a phononic system is not pairwise.~\citep{Rodin2019} In other words, the total energy does not equal the sum of interaction energies between pairs of defects in a way that one finds in, say, electrostatic charge interaction. The novel ability to work with multiple phonon branches allows one to study polyatomic systems. As was suggested in Ref.~\citep{Pavlov2019}, polyatomic lattices with high Debye frequencies, currently used in the studies of superconductivity, could be suitable candidates for observing PCE in two and three dimensions.

The paper is organized as follows. In Sec.~\ref{sec:General_Formalism}, we present the derivation of this work's main result and discuss how it relates to exact diagonalization in Sec.~\ref{sec:Exact_Diagonalization}. Section~\ref{sec:Diatomic_Molecule} is dedicated to the simplest systems treatable by our formalism: diatomic molecules. We adapt our formalism to infinitely large periodic systems, where exact diagonalization fails, in Sec.~\ref{sec:Periodic_Systems}. To demonstrate our formalism in action, Sec.~\ref{sec:1D_chain} is dedicated to the study of infinite one-dimensional chains that can be compared to large-but-finite systems treated using exact diagonalization. Concluding remarks are given in Sec.~\ref{sec:Conclusions}.

\section{General Formalism}
\label{sec:General_Formalism}

We begin our discussion by constructing a framework to handle systems of any dimensionality with an arbitrary defect number and arrangement. To make the derivation as transparent as possible, we approach it systematically by first providing a second-quantized Hamiltonian for such a general system. Next, we translate this Hamiltonian into the imaginary-time action to calculate the system's free energy and, consequently, the defect interaction energy.

\subsection{Hamiltonian}
\label{sec:Hamiltonian}

Even though PCE is typically formulated for crystalline materials which support phonon modes, it is more general and, in fact, easier to derive it for an arbitrary system with vibrational modes without insisting on crystal symmetry. A general Hamiltonian operator for a $D$-dimensional system of this type can be written as
\begin{align}
    \hat{H} &= 
    \frac{1}{2}\sum_{j}
    \frac{ \hat{\mathbf{p}}_j^\dagger\hat{\mathbf{p}}_j}{m_j}
    +
    \frac{1}{2}\sum_{jk}
    \hat{\mathbf{u}}_{j}^\dagger V_{jk}\hat{\mathbf{u}}_{k}
    \nonumber
    \\
    &+
    \frac{1}{2}\sum_{jk}
    \hat{\mathbf{p}}_{j}^\dagger 
    \Lambda_{jk}
    \hat{\mathbf{p}}_{k}
    +
    \frac{1}{2}\sum_{jk}
    \hat{\mathbf{u}}_{j}^\dagger 
    \Delta_{jk}
    \hat{\mathbf{u}}_{k}
    \label{eqn:H}
    \,,
\end{align}
where $\hat{\mathbf{p}}_j$ and $\hat{\mathbf{u}}_j$ are momentum and displacement operator vectors of length $D$ for the $j$th atom, respectively. The first line describes the system in the absence of defects using the harmonic approximation: $m_j$ is the mass of the $j$th atom and $V_{jk}$ is a $D\times D$ harmonic coupling matrix between the displacements of the $j$th and $k$th atoms. $\Delta_{jk}$ and $\Lambda_{jk}$ in the second line are symmetric $D\times D$ matrices and correspond to defects. The former can be used to describe a change in the force constant between atoms or a local external potential, while the latter can be used to represent effects like the change in the atomic mass, in which case $\Lambda_{jj} = \mathbf{1}_{D\times D}\otimes \left(M_j^{-1} - m_j^{-1}\right)$, where $M_j$ is the new mass. Note that the sums in the second line include all atoms in the system, even if the corresponding $\Lambda_{jk}$ and $\Delta_{jk}$ are zero.

To translate the problem into the language of second quantization, we write the position and momentum operators as
\begin{align}
	\hat{\mathbf{u}}_l
    &=
	\frac{1}{\sqrt{2}}\sum_{s}
	\left(b_{s}
    +
    b_{s}^\dagger \right)
    \overbrace{\sqrt{\frac{1}{m_l \Omega_s} }
    \boldsymbol{\varepsilon}_{s,l}}^{U_{s,l}}
    \,,
    \label{eqn:u}
	\\
	\hat{\mathbf{p}}_l
	&=
	\frac{i}{\sqrt{2}}\sum_{s}
	\left(-b_{s}
    +
    b_{s}^\dagger\right)
	\underbrace{\sqrt{	m_l\Omega_{s}}
	\boldsymbol{\varepsilon}_{s,l}}_{P_{s,l}}
	\,.
	\label{eqn:p}
\end{align}
Here $\boldsymbol{\varepsilon}_s$ is the normalized mode eigenvector containing the amplitudes for all the atoms in the solid. It is obtained by solving $\Omega_s^2 \boldsymbol{\varepsilon}_s = \mathbf{m}^{-1/2}\mathbf{V}\mathbf{m}^{-1/2}\boldsymbol{\varepsilon}_s$, where $\mathbf{V}$ is the matrix of force constants and $\mathbf{m}$ is a block-diagonal matrix of $m_j \mathbf{1}_{D\times D}$. $\boldsymbol{\varepsilon}_{s,l}$ is the segment of this eigenvector of length $D$ corresponding to the $l$th atom and $b_{s}$ ($b^\dagger_{s}$) are bosonic annihilation (creation) operators for the vibrational mode $s$.

Plugging Eqs.~\eqref{eqn:u} and \eqref{eqn:p} into the first two terms of Eq.~\eqref{eqn:H} yields the familiar harmonic mode Hamiltonian
\begin{equation}
	\hat{H}_0 = \sum_{s}
	\Omega_{s}
	\left(
    b_{s}^\dagger
	b_{s} + \frac{1}{2}
    \right)\,.
    \label{eqn:H0}
\end{equation}

For the defect part, we write
\begin{align}
    \hat{H}_\mathrm{def} &=\frac{1}{4}
    \sum_{ss'}
	\mathbf{P}_{s'}^T
    \Lambda\mathbf{P}_{s}
	\big(-b_{s'}^\dagger
    +
    b_{s'}\big)
	\big(-b_{s}
    +
    b_{s}^\dagger\big)
    \nonumber
    \\
    &
    +
    \frac{1}{4}
    \sum_{ss'}
	\mathbf{U}_{s'}^T
    \Delta
    \mathbf{U}_{s}
	\big(b_{s'}
    +
    b_{s'}^\dagger \big)
	\big(b_{s}
    +
    b_{s}^\dagger \big)\,,
    \label{eqn:H_D}
\end{align}
where $\mathbf{P}_s = \Omega_s^{1/2}\mathbf{m}^{1/2}\boldsymbol{\varepsilon}_s$ ($\mathbf{U}_s= \Omega_s^{-1/2}\mathbf{m}^{-1/2}\boldsymbol{\varepsilon}_s$) is the column vector of $P_{s,l}$ ($U_{s,l}$) and $\Lambda$ ($\Delta$) is the matrix of $\Lambda_{jk}$ ($\Delta_{jk}$).

Whereas Eq.~\eqref{eqn:H0} is normal-ordered, Eq.~\eqref{eqn:H_D} is not. Therefore, we commute the operators to establish the ordering necessary for the application of the path integral formalism:
\begin{align}
    &\big(\pm b_{s'}^\dagger
    +
    b_{s'} \big)
	\big(\pm b_{s}
    +
    b_{s}^\dagger \big)
    \nonumber
    \\
    =& 
    b_{s'}^\dagger b_{s}
    +b_{s}^\dagger b_{s'} 
    \pm  b_{s'} b_{s}
    \pm b_{s'}^\dagger b_{s}^\dagger
    + \delta_{ss'}\,. 
    \label{eqn:Normal_Order}
\end{align}
As expected, commuting the operators produces constant energy terms. Combining these terms with the vacuum energy portion of Eq.~\eqref{eqn:H0} yields
\begin{align}
	F_0 &= \sum_{s} \frac{\Omega_{s}}{2} 
	+
	\frac{1}{4}
    \sum_{s}\mathbf{P}_s^T\Lambda\mathbf{P}_s
	+
	\frac{1}{4}
    \sum_{s}\mathbf{U}_s^T\Delta\mathbf{U}_s\,.
	\label{eqn:F0}
\end{align}
Note that if $\Delta$ and $\Lambda$ are block-diagonal in defects (i.e., $\Lambda_{ij} = \Lambda_{ij}\delta_{ij}$, same for $\Delta$), $F_0$ contains no defect-defect cross-terms and, therefore, no interaction between the defects.

$F_0$ in Eq.~\eqref{eqn:F0}, originating from the commutation of the creation and annihilation operators, is a consequence of the Heisenberg uncertainty principle. For a pristine system ($\Delta = \Lambda = 0$), it is the only contribution to the Helmholtz free energy at zero temperature. However, if the system contains defects, there is another term that adds to the free energy even at zero temperature. It is precisely that term that gives rise to PCE if $\Delta$ and $\Lambda$ are block-diagonal, as we will show below.

\subsection{Action and Partition Function}
\label{sec:Action}

The normal-ordered operator-dependent part of the Hamiltonian can be straightforwardly transcribed into the imaginary-time action
\begin{widetext}
\begin{align}
    S &= \sum_{n}
    \bigg\{\sum_{s}
	\bar{\phi}_{n,s}\left(
    -i\omega_n + \Omega_{s}
    \right)
    \phi_{n,s}
    \nonumber
    \\
    &+
    \frac{1}{4}
    \sum_{ss'}
	\mathbf{P}_{s'}^T\Lambda\mathbf{P}_s
    \left( 
    \bar{\phi}_{n,s'} \phi_{n,s}
    +
    \bar{\phi}_{n,s} \phi_{n,s'} 
    -  \phi_{-n,s'} \phi_{n,s}
    - \bar{\phi}_{n,s'} \bar{\phi}_{-n,s}\right)
    \nonumber
    \\
 &+
   \frac{1}{4}
    \sum_{ss'}
	\mathbf{U}_{s'}^T\Delta\mathbf{U}_s
    \left( 
    \bar{\phi}_{n,s'} \phi_{n,s}
    +
    \bar{\phi}_{n,s} \phi_{n,s'} 
    +  \phi_{-n,s'} \phi_{n,s}
    + \bar{\phi}_{n,s'} \bar{\phi}_{-n,s}\right)\bigg\}\,,
    \label{eqn:S_Matsubara}
\end{align}
\end{widetext}
where $\omega_n$ are bosonic Matsubara frequencies. Note that while the products of the fields corresponding to annihilation and creation operators carry the same Matsubara frequency, products of two creation/annihilation fields have the opposite frequency index.

Exponentiating $-S$ and integrating over all fields gives the partition function, from which the Helmholtz free energy can be obtained. Before performing the field integrals, however, we note that, for non-zero Matsubara components, the fields in the defect portion of the action enter either as symmetric or antisymmetric in $n$, which is easier to see if we write
\begin{align}
	&S^{n\neq0} = \sum_{n,s}\bar\phi_{n,s}\left(-i\omega_n + \Omega_{s}\right)\phi_{n,s}
	\nonumber
	\\
    +&
     \frac{1}{4}
    \sum_{n,ss'}\mathbf{P}_{s'}^T\Lambda\mathbf{P}_s
    \left(
    \bar\phi_{n,s'} -
    \phi_{-n,s'}\right)
     \left(\phi_{n,s}
    -
    \bar\phi_{-n,s}\right)\,,
    \nonumber
    \\
    +&
    \frac{1}{4}
    \sum_{n,ss'}
\mathbf{U}_{s'}^T\Delta\mathbf{U}_s
    \left(\bar\phi_{n,s'}+\phi_{-n,s'} \right)
    \left(\phi_{n,s}
    +
    \bar\phi_{-n,s}\right)\,.
    \label{eqn:S_phi_n>0}
\end{align}
Therefore, it is useful to introduce a change of variables $\psi_{n,s}^\pm = \left(\phi_{n,s} \pm \bar\phi_{-n,s}\right)/\sqrt{2}$. Because $\bar\psi^{\pm}_{n,s} = \pm \psi^{\pm}_{-n,s}$, only $\psi^\pm_{n > 0}$ are unique. Explicitly, the momentum perturbation term becomes
\begin{align}
     &\sum_{n>0,ss'}
    \frac{\mathbf{P}_{s'}^T\Lambda\mathbf{P}_s}{2}
	\left(
	\bar\psi_{n,s'}^- \psi_{n,s}^-
	+
	\bar\psi_{-n,s'}^- \psi_{-n,s}^-
	\right)
	\nonumber
	\\
	=&\sum_{n>0,ss'}
    \frac{\mathbf{P}_{s'}^T\Lambda\mathbf{P}_s}{2}
	\left(
	\bar\psi_{n,s'}^- \psi_{n,s}^-
	+
	\psi_{n,s'}^- \bar\psi_{n,s}^-
	\right)
	\nonumber
	\\
	=&\sum_{n>0,ss'}
    \mathbf{P}_{s'}^T\Lambda\mathbf{P}_s
	\bar\psi_{n,s'}^- \psi_{n,s}^-\,.
\end{align}
For the last step, we use the fact that the two terms in the parentheses are related by the interchange of $s\leftrightarrow s'$. The prefactor scalar remains invariant under this interchange because $\Lambda$ is symmetric [$\mathbf{P}_{s'}^T\Lambda\mathbf{P}_s = \left(\mathbf{P}_{s'}^T\Lambda\mathbf{P}_s\right)^T
= \mathbf{P}_{s}^T\Lambda\mathbf{P}_{s'}$], allowing us to combine the terms in the parentheses. A similar procedure can be performed for the $\mathbf{U}$-term in Eq.~\eqref{eqn:S_phi_n>0}, leading to
\begin{align}
	S^{n>0}_\psi &=
	\sum_{n>0,ss'} 
	\bar\Psi_{n,s'}
	\Bigg[
	 \overbrace{\begin{pmatrix}
		\Omega_{s} &-i\omega_n 
		\\
		-i\omega_n & \Omega_{s} 
	\end{pmatrix}}^{-\Gamma^{-1}_{n,s}} \delta_{ss'}
	\nonumber
	\\
	&+
	\underbrace{\begin{pmatrix}
    \mathbf{U}_{s'}^T
	&0
	\\
	0&
	\mathbf{P}_{s'}^T
    \end{pmatrix}}_{R_{s'}^T}
    \begin{pmatrix}
    \Delta
	&0
	\\
	0&
    \Lambda
    \end{pmatrix}
    \underbrace{
    \begin{pmatrix}
	\mathbf{U}_{s}
	&0
	\\
	0&
	\mathbf{P}_{s}
    \end{pmatrix}}_{R_{s}}
    \Bigg]
	\Psi_{n,s}
	\,,
	\label{eqn:S_psi}
\end{align}
where $\bar\Psi_{n,s'} = \begin{pmatrix} \bar\psi^+_{n,s'} & \bar\psi^-_{n,s'} \end{pmatrix}$ and $\Gamma^{-1}_{n,s}$ originates from the first line of Eq.~\eqref{eqn:S_phi_n>0}.

Taking the Gaussian field integral for each $n$ yields the partition function
\begin{align}
    \mathcal{Z} &=\mathcal{Z}_0\prod_{n>0}
    \left|\beta\left(-\Gamma_n^{-1} + R^T
	\begin{pmatrix}
		\Delta&0
		\\
		0&\Lambda
	\end{pmatrix}
	R\right)\right|^{-1}
	\label{eqn:Z}
\end{align}
where $R^T$ is a column vector of $R_{s}^T$ and $\Gamma_n^{-1}$ is a block-diagonal matrix of $\Gamma_{n,s}^{-1}$. We will address $\mathcal{Z}_0$, coming from the $n = 0$ portion of the action shortly.  For a later convenience, we rewrite the determinant term as
\begin{equation}
    \mathcal{Z}_{n\neq 0} =
	\left|-\beta\Gamma_n^{-1} \right|^{-1}
	\left|1+\Xi(i\omega_n)
	\begin{pmatrix}
		\Delta &0
		\\
		0&\Lambda
	\end{pmatrix}
	\right|^{-1}\,,
	\label{eqn:Z_split}
\end{equation}
where $\Xi(i\omega_n) =-R\Gamma_n R^T$. Explicitly,
\begin{align}
   & \Xi(i\omega_n) =
	\sum_{s}
	 R_s
	\begin{pmatrix}
		\frac{\Omega_{s}}{\omega_n^2 + \Omega_{s}^2} 
		&
		\frac{i\omega_n}{\omega_n^2 + \Omega_{s}^2}
		\\
		\frac{i\omega_n}{\omega_n^2 + \Omega_{s}^2} 
		& 
		\frac{\Omega_{s}}{\omega_n^2 + \Omega_{s}^2}
	\end{pmatrix}
	R_s^T
	\nonumber
	\\
	=&
	\sum_{s}
	\begin{pmatrix} \mathbf{m}^{-\frac{1}{2}}\boldsymbol{\varepsilon}_{s}\boldsymbol{\varepsilon}_{s}^T\mathbf{m}^{-\frac{1}{2}}
    &
    i\omega_n  \mathbf{m}^{-\frac{1}{2}}\boldsymbol{\varepsilon}_{s}\boldsymbol{\varepsilon}_{s}^T\mathbf{m}^{\frac{1}{2}}
    \\
    i\omega_n  \mathbf{m}^{\frac{1}{2}}\boldsymbol{\varepsilon}_{s}\boldsymbol{\varepsilon}_{s}^T\mathbf{m}^{-\frac{1}{2}}
    &
     \Omega_{s}^2 \mathbf{m}^{\frac{1}{2}}\boldsymbol{\varepsilon}_{s}\boldsymbol{\varepsilon}_{s}^T\mathbf{m}^{\frac{1}{2}}
	\end{pmatrix}\frac{1}{\omega_n^2 + \Omega_{s}^2}
	\nonumber
	\\
	=&
	\begin{pmatrix}
		\mathbf{m}^{-\frac{1}{2}}
		\\
		 i\omega_n\mathbf{m}^{\frac{1}{2}}
	\end{pmatrix}
	\Pi(i\omega_n)
	\begin{pmatrix}
		\mathbf{m}^{-\frac{1}{2}}&i\omega_n\mathbf{m}^{\frac{1}{2}}
	\end{pmatrix}
	+
	\begin{pmatrix}
		0&0
		\\
		0 & \mathbf{m}
	\end{pmatrix}
	\label{eqn:Xi}
\end{align}
with
\begin{align}
    \left[\Pi(z)\right]_{ij}&=
   \sum_{s}\frac{ \boldsymbol{\varepsilon}_{s,i} \otimes \boldsymbol{\varepsilon}_{s,j} }{-z^2+\Omega^2_{s}}\,,
    \label{eqn:Pi}
\end{align}
as $D\times D$ matrix. The last equality in Eq.~\eqref{eqn:Xi} relies on the fact that $\sum_s \boldsymbol{\varepsilon}_{s}\boldsymbol{\varepsilon}_{s}^T = 1$ due to the orthonormality of the eigenvectors.

So far, we have addressed the action corresponding non-zero Matsubara frequencies. The remaining term in Eq.~\eqref{eqn:S_Matsubara} can be written as
\begin{align}
    S^{n=0} 
    &= 
    \frac{1}{2}\sum_{ss'}
    \bar\Phi_{0,s'}
    \Bigg[
    \overbrace{\begin{pmatrix}
		\Omega_{s} &0
		\\
		0 & \Omega_{s} 
	\end{pmatrix}}^{-\Gamma^{-1}_{0,s}}
     \delta_{ss}
     \nonumber
     \\
    &+
    \underbrace{\begin{pmatrix}
        \mathbf{U}_{s'}^T 
         &
         \mathbf{P}_{s'}^T
         \\
         \mathbf{U}_{s'}^T 
         &
         -\mathbf{P}_{s'}^T 
    \end{pmatrix}}_{\tilde{R}^T_{s'}}
    \begin{pmatrix}
    \frac{\Delta }{2}&0
    \\
    0 & \frac{\Lambda }{2}
    \end{pmatrix}
    \underbrace{\begin{pmatrix}
         \mathbf{U}_{s}
         &
     \mathbf{U}_{s}
          \\
        \mathbf{P}_{s}
         &
          -\mathbf{P}_{s}
    \end{pmatrix}}_{\tilde{R}_{s}}
    \Bigg]\Phi_{0,s}\,,
\end{align}
where $\begin{pmatrix}
        \bar{\phi}_{0,s} & \phi_{0,s}
    \end{pmatrix}= \bar\Phi_{0,s}$. Integrating over $\Phi_{0}$ gives
\begin{align}
    \mathcal{Z}_0 &= \left|\beta\left(-\Gamma^{-1}_0 + 
    \tilde{R}^T
    \begin{pmatrix}
    \frac{\Lambda }{2}&0
    \\
    0 & \frac{\Delta }{2}
    \end{pmatrix}
    \tilde{R}\right)\right|^{-\frac{1}{2}}
    \nonumber
    \\
    &= \left|-\beta\Gamma^{-1}_0 \right|^{-\frac{1}{2}}
    \left|1 - 
     \frac{\tilde{R}
     \Gamma_0
    \tilde{R}^T}{2}
    \begin{pmatrix}
    \Delta&0
    \\
    0 & \Lambda
    \end{pmatrix}
   \right|^{-\frac{1}{2}}\,.
   \label{eqn:Z_0}
\end{align}
Explicitly,
\begin{align}
    - 
     \frac{\tilde{R}
     \Gamma_0
    \tilde{R}^T}{2}
    &=\frac{1}{2}
    \sum_{s}
    \frac{1}{\Omega_{s}}
    \begin{pmatrix}
        \mathbf{U}_{s}
         &
         \mathbf{U}_{s}
          \\
        \mathbf{P}_{s}
         &
          - \mathbf{P}_{s}
    \end{pmatrix}
    \begin{pmatrix}
        \mathbf{U}_{s}^T 
         &
         \mathbf{P}_{s}^T 
         \\
         \mathbf{U}_{s}^T
         &
         -\mathbf{P}_{s}^T 
    \end{pmatrix}
    \nonumber
    \\
     &=\begin{pmatrix}
	\mathbf{m}^{-\frac{1}{2}}
	\Pi(0)
	\mathbf{m}^{-\frac{1}{2}}
    &
    0
    \\
    0
    &
   \mathbf{m}
	\end{pmatrix}
     = \Xi(0)\,.
\end{align}

\subsection{Free Energy}
\label{sec:Free_Energy}

The free energy is obtained from the partition function using $F = - T\ln\mathcal{Z}$, where $T$ is the temperature. Combining Eqs.~\eqref{eqn:F0}, \eqref{eqn:Z}, \eqref{eqn:Z_split} and ~\eqref{eqn:Z_0}  yields
\begin{align}
    F &= \sum_{s}\frac{\Omega_s}{2} + T\sum_{n\geq 0} \frac{1}{2^{\delta_{n,0}}}\ln\left|-\beta\Gamma_n^{-1}\right|
     + F_\mathrm{def}\,,
	\label{eqn:F_Full}
	\\
	F_\mathrm{def} &= \frac{1}{4}
    \sum_{s}\mathbf{P}_s^T\Lambda\mathbf{P}_s
	+
	\frac{1}{4}
    \sum_{s}\mathbf{U}_s^T\Delta\mathbf{U}_s
    \nonumber
   \\
    &+
    T\sum_{n \geq 0}\frac{1}{2^{\delta_{n,0}}}\ln \left|1+\Xi(i\omega_n)
	\begin{pmatrix}
		\Delta &0
		\\
		0&\Lambda
	\end{pmatrix}
	\right|\,.
	\label{eqn:F_def}
\end{align}
Here, $F_\mathrm{def}$ is the defect-generated part of the free energy and the rest of the terms in Eq.~\eqref{eqn:F_Full} constitute the free energy of the pristine system.

As we stated earlier, $\Lambda$ and $\Delta$ matrices, as well as $\mathbf{m}$ and $\Pi$ in Eq.~\eqref{eqn:Xi}, include all the atoms in the system. It is clear from Eq.~\eqref{eqn:F_def}, however, that the atoms not subject to a $\Delta$ or $\Lambda$ perturbation do not contribute to the free energy as the corresponding entries in the $\Xi$ matrix get multiplied by zero. Therefore, when computing $\Xi$, we only need to retain the perturbed atoms.

To calculate the interaction energy $F_I$ for a collection of defects, we subtract the $F_\mathrm{def}$ for each individual defect from the total multi-defect $F_\mathrm{def}$. If $\Delta$ and $\Lambda$ are block-diagonal, the terms in the first line of Eq.~\eqref{eqn:F_def} cancel to give
\begin{align}
    F_I &= T\sum_{n \geq 0}\frac{1}{2^{\delta_{n,0}}}\ln \left|1+\Xi(i\omega_n)
	\begin{pmatrix}
		\Delta &0
		\\
		0&\Lambda
	\end{pmatrix}
	\right|
	\nonumber
   \\
    &-
    T\sum_{n \geq 0}\frac{1}{2^{\delta_{n,0}}}\ln \left|1+\Xi_\mathrm{diag}(i\omega_n)
	\begin{pmatrix}
		\Delta &0
		\\
		0&\Lambda
	\end{pmatrix}
	\right|\,,
	\label{eqn:F_I}
\end{align}
where $\Xi_\mathrm{diag}$ is a variant of $\Xi$ constructed using $\Pi_\mathrm{diag}$ which itself contains only the diagonal blocks of $\Pi$. In other words, $\Pi_\mathrm{diag}$ removes the coupling between different atoms to give the individual $F_\mathrm{def}$'s.

It is useful to separate the $n = 0$ term in Eq.~\eqref{eqn:F_I}:
\begin{align}
   &\ln \left|1+\begin{pmatrix}
	\mathbf{m}^{-\frac{1}{2}}
	\Pi(0)
	\mathbf{m}^{-\frac{1}{2}}
    &
    0
    \\
    0
    &
   \mathbf{m}
	\end{pmatrix}
	\begin{pmatrix}
		\Delta &0
		\\
		0&\Lambda
	\end{pmatrix}
	\right|
	\nonumber
	\\
	-&\ln \left|1+\begin{pmatrix}
	\mathbf{m}^{-\frac{1}{2}}
	\Pi_\mathrm{diag}(0)
	\mathbf{m}^{-\frac{1}{2}}
    &
    0
    \\
    0
    &
   \mathbf{m}
	\end{pmatrix}
	\begin{pmatrix}
		\Delta &0
		\\
		0&\Lambda
	\end{pmatrix}
	\right|
	\nonumber
	\\
	=&
	   \ln \left|1+\mathbf{m}^{-\frac{1}{2}}
	\Pi(0)
	\mathbf{m}^{-\frac{1}{2}}\Delta
	\right|
	\nonumber
	\\
	-&
	   \ln \left|1+\mathbf{m}^{-\frac{1}{2}}
	\Pi_\mathrm{diag}(0)
	\mathbf{m}^{-\frac{1}{2}}\Delta
	\right|
\end{align}
so that the final form for the interaction energy becomes
\begin{align}
    F_I &= T\sum_{n > 0}\ln \left|1+\Xi(i\omega_n)
	\begin{pmatrix}
		\Delta &0
		\\
		0&\Lambda
	\end{pmatrix}
	\right|
	\nonumber
   \\
    &-
    T\sum_{n > 0}\ln \left|1+\Xi_\mathrm{diag}(i\omega_n)
	\begin{pmatrix}
		\Delta &0
		\\
		0&\Lambda
	\end{pmatrix}
	\right|
	\nonumber
	\\
	&+ \frac{T}{2}\ln \left|1+\mathbf{m}^{-\frac{1}{2}}
	\Pi(0)
	\mathbf{m}^{-\frac{1}{2}}\Delta
	\right|
	\nonumber
	\\
	&-
	   \frac{T}{2}\ln \left|1+\mathbf{m}^{-\frac{1}{2}}
	\Pi_\mathrm{diag}(0)
	\mathbf{m}^{-\frac{1}{2}}\Delta
	\right|\,.
	\label{eqn:F_I_Final}
\end{align}

Before moving to concrete examples, we provide a summary of the steps that one takes to calculate $F_I$ for a general system with diagonal $\Delta$ and $\Lambda$ which will be the focus of the rest of this paper:
\begin{enumerate}
    \item Determine the masses of the system atoms hosting the defects and construct $\mathbf{m}^{\pm1/2}$.
    \item Construct $\Delta$ and $\Lambda$, each of the same dimension as $\mathbf{m}^{\pm 1/2}$. If a particular mass only has the $\Delta$-type or the $\Lambda$-type perturbation, the corresponding elements in the other matrix will be zero.
    \item Compute the mode frequencies $\Omega_{s}$ and the corresponding vectors $\boldsymbol{\varepsilon}_{s}$ for the host system.
    \item Calculate the $\Pi(z)$ matrix and use it to construct $\Xi(i\omega_n)$ and $\Xi_\mathrm{diag}(i\omega_n)$.
    \item Plug the $\Pi(z)$ and $\Xi(z)$ matrices into Eq.~\eqref{eqn:F_I_Final} and perform the summation over the Matsubara frequencies.
\end{enumerate}

\section{Exact Diagonalization}
\label{sec:Exact_Diagonalization}

For systems that are not prohibitively large, it is possible to validate the path integral results using exact diagonalization. Recall that the free energy of non-interacting Bose gas is given by
\begin{equation}
    F = \sum_s\frac{\Omega_s}{2} + T\ln\left(1 - e^{-\Omega_s / T}\right)\,,
    \label{eqn:F_Bose}
\end{equation}
where $\Omega_s$ are the energies of the bosonic states (mode frequencies in the context of this work). As in Sec.~\ref{sec:General_Formalism}, the interaction energy between defects is calculated by first subtracting the pristine-system $F$ from the multi-defect $F$ and also from single-defect $F$'s for each individual defect to obtain the corresponding $F_\mathrm{def}$'s. Then, by subtracting the single-defect $F_\mathrm{def}$'s from the multi-defect one, we obtain $F_I$. We will compare the exact diagonalization results with our formalism in the subsequent sections when we explore concrete examples. At this point, however, it is useful to consider the high-$T$ limit of $F_I$ as obtained from Eq.~\eqref{eqn:F_Bose}.

First, note that as $T\rightarrow \infty$
\begin{equation}
    F \approx \sum_s T\ln\left(\frac{\Omega_s}{T}\right) = 
    T\ln\left(\prod_s\frac{\Omega_s}{T}\right)
    =
    T\ln\left|\frac{\boldsymbol{\Omega}}{T}\right|
    \,,
    \label{eqn:F_Bose_High_T}
\end{equation}
where $\boldsymbol{\Omega}$ is a diagonal matrix of $\Omega_s$ obtained by the orthogonal transformation $\boldsymbol{\Omega}^2 = O^T\mathbf{m}^{-\frac{1}{2}}\mathbf{V}\mathbf{m}^{-\frac{1}{2}}O$. Next, using $\ln|\boldsymbol{\Omega}/T| = \ln|\boldsymbol{\Omega}^2/T^2|/2$, we get
\begin{equation}
    F =  \frac{T}{2}\ln\left|\frac{O^T\mathbf{m}^{-\frac{1}{2}}\mathbf{V}\mathbf{m}^{-\frac{1}{2}}O}{T^2}\right|
    =  \frac{T}{2}\ln\left|\frac{\mathbf{m}^{-1}\mathbf{V}}{T^2}\right|\,,
\end{equation}
leading to
\begin{align}
    F_\mathrm{def} &= \frac{T}{2}\ln\left|\frac{\mathbf{m}^{-1}\mathbf{V}}{T^2}\right|-\frac{T}{2}\ln\left|\frac{\mathbf{m}_0^{-1}\mathbf{V}_0}{T^2}\right|
    \nonumber
    \\
    &= \frac{T}{2}\ln\left|\mathbf{m}^{-1}\mathbf{m}_0\right|-\frac{T}{2}\ln\left|\mathbf{V}^{-1}\mathbf{V}_0\right|\,,
\end{align}
where the matrices $\mathbf{m}$ and $\mathbf{V}$ include the defects, while $\mathbf{m}_0$ and $\mathbf{V}_0$ are their unperturbed counterparts. Because $\mathbf{m}$ and $\mathbf{m}_0$ are diagonal, the first term in the expression above can be written as $DT / 2 \sum_j \ln(m_0^j / m^j)$ where the sum runs over all the perturbed atoms. It is easy to see that when we subtract the single-defect $F_\mathrm{def}$'s from the multiple-defect $F_\mathrm{def}$, this term cancels: $DT / 2 \sum_j \ln(m_0^j / m^j) -  \sum_j \left[DT / 2\ln(m_0^j / m^j)\right]$. The remaining part gives
\begin{align}
    F_I &=
    -\frac{T}{2}\ln\left|\mathbf{V}_\mathrm{all}^{-1} \mathbf{V}_0\right| + \frac{T}{2}\sum_j\ln\left|\mathbf{V}_j^{-1}\mathbf{V}_0\right|
    \nonumber
    \\
    &=
    \frac{T}{2}\ln\left|1+\mathbf{V}_0^{-1}\Delta_\mathrm{all} \right| 
    -
    \frac{T}{2}\sum_j\ln\left|1+\mathbf{V}_0^{-1}\Delta_j \right|\,,
    \label{eqn:F_I_Exact_Diagonalization}
\end{align}
where $\mathbf{V}_\mathrm{all} = \mathbf{V}_0 + \Delta_\mathrm{all}$ is the force constant matrix with all the perturbations, and $\mathbf{V}_j = \mathbf{V}_0 + \Delta_j$ is the matrix with a single perturbation on the $j$th atom.

At this point, we drop the subscript $0$ from $\mathbf{V}_0$ as we have separated the perturbation in Eq.~\eqref{eqn:F_I_Exact_Diagonalization}, making the subscript redundant. Using the orthogonal transformation between $\mathbf{V}$ and $\mathbf{\Omega}^2$, we can write $\boldsymbol{\Omega}^{-2} = O^T\mathbf{m}^{\frac{1}{2}}\mathbf{V}^{-1}\mathbf{m}^{\frac{1}{2}}O\rightarrow \mathbf{m}^{-\frac{1}{2}}O\boldsymbol{\Omega}^{-2}O^T\mathbf{m}^{-\frac{1}{2}} = \mathbf{V}^{-1}$. The orthogonal matrix $O$ is a row vector of $\boldsymbol{\varepsilon}$'s so that $O\boldsymbol{\Omega}^{-2}O^T = \sum_s \boldsymbol{\varepsilon}_s\boldsymbol{\varepsilon}^T_s / \Omega_s^2 = \Pi(0)$ and $\mathbf{m}^{-\frac{1}{2}} \Pi(0)\mathbf{m}^{-\frac{1}{2}} = \mathbf{V}^{-1}$. Plugging this into Eq.~\eqref{eqn:F_I_Exact_Diagonalization} and making use of $\Pi_\mathrm{diag}$ to combine the $j$ summation gives the last two terms of Eq.~\eqref{eqn:F_I_Final}. By demonstrating the equality between the high-$T$ result obtained from exact diagonalization and the zeroth Matsubara term, we have confirmed that at high-temperature $F_I$ is dominated by this term, as expected.

Let us now address the significance of the $n = 0$ term. Recall that the Helmholtz free energy is defined as $F \equiv E - TS$, where $E$ is the internal energy of the system and $S$ is the entropy. We can write $F_I = E_I - T S_I$, where $E_I$ is the ``internal interaction energy" and $S_I$ is the entropy difference between the many-defect configuration and single-defect systems. By performing the high-$T$ analysis, we managed to isolate the term that is proportional to $T$, making it a candidate for $S_I$. We identify this term in Eq.~\eqref{eqn:F_I_Final} as the only portion of $F_I$ that grows linearly with $T$ (all other terms are nonlinear in $T$ because $\omega_n \propto T$), confirming its identity as the entropy. As a result, to calculate the internal interaction energy $E_I$ using our formalism instead of the Helmholtz free energy, one simply needs to drop the zeroth Matsubara term from Eq.~\eqref{eqn:F_I_Final}. To get the same from exact diagonalization, one subtracts Eq.~\eqref{eqn:F_I_Exact_Diagonalization} from Eq.~\eqref{eqn:F_Bose}.

\section{Diatomic Molecule}
\label{sec:Diatomic_Molecule}

\begin{figure}
    \centering
    \includegraphics[width = \columnwidth]{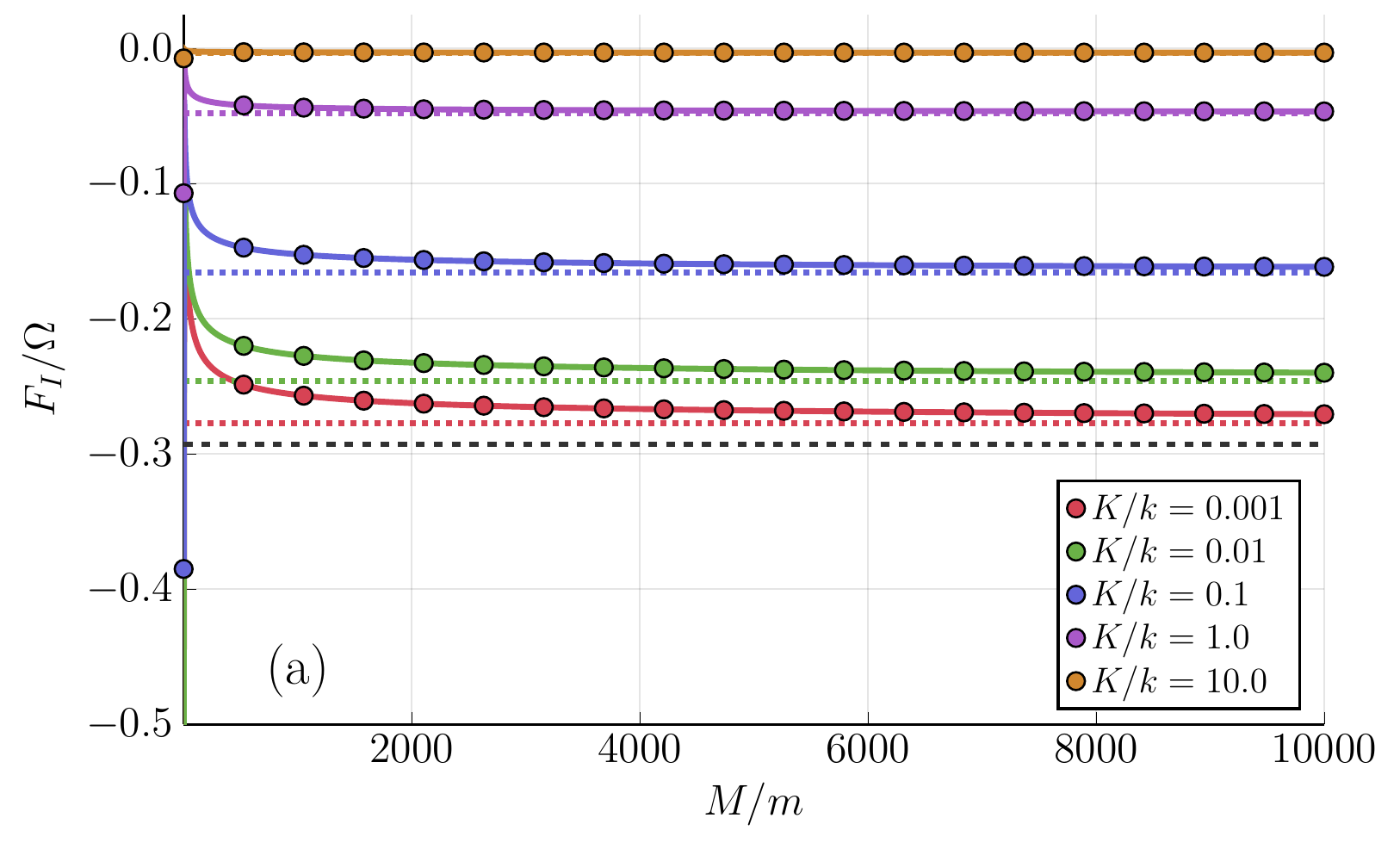}
    \includegraphics[width = \columnwidth]{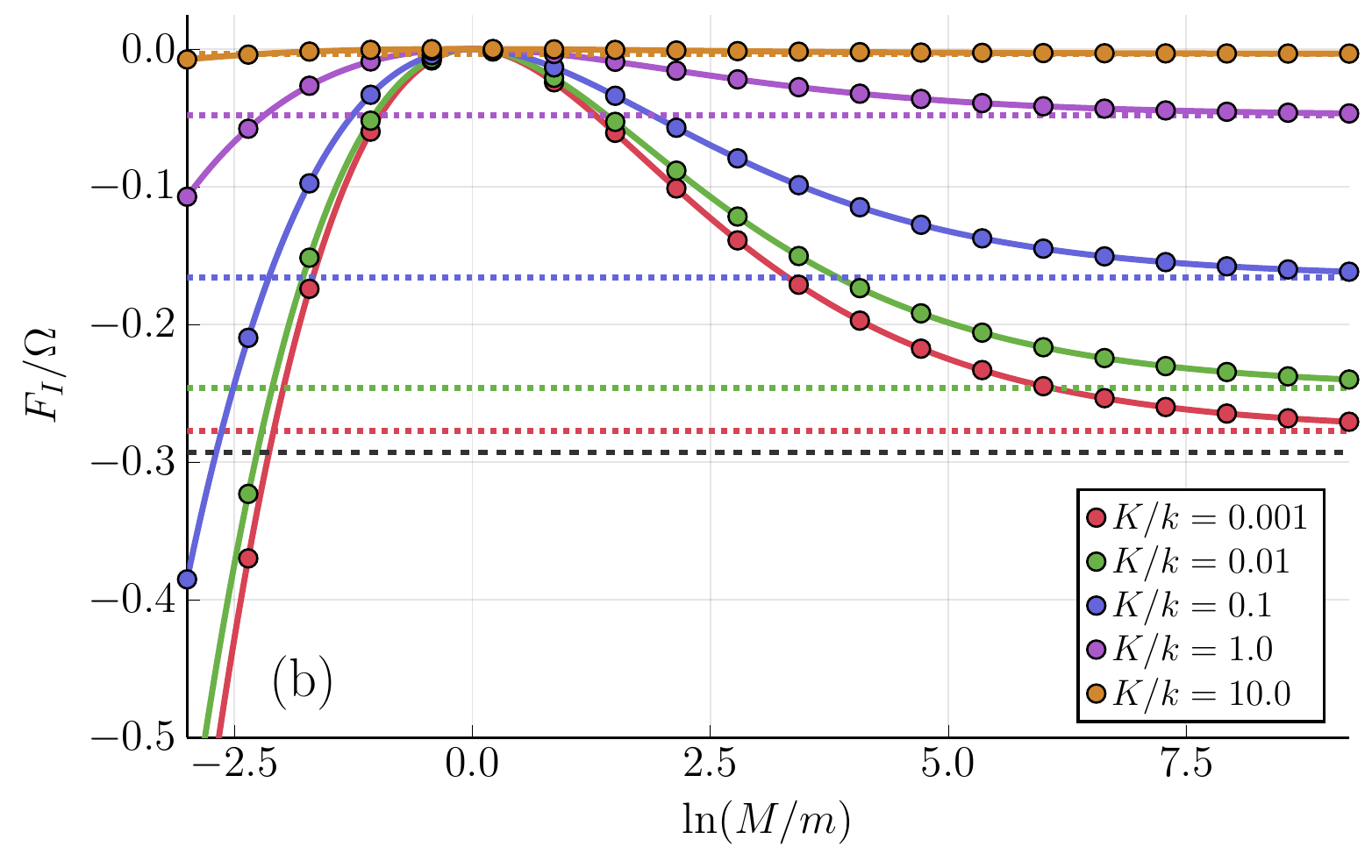}
    \includegraphics[width = \columnwidth]{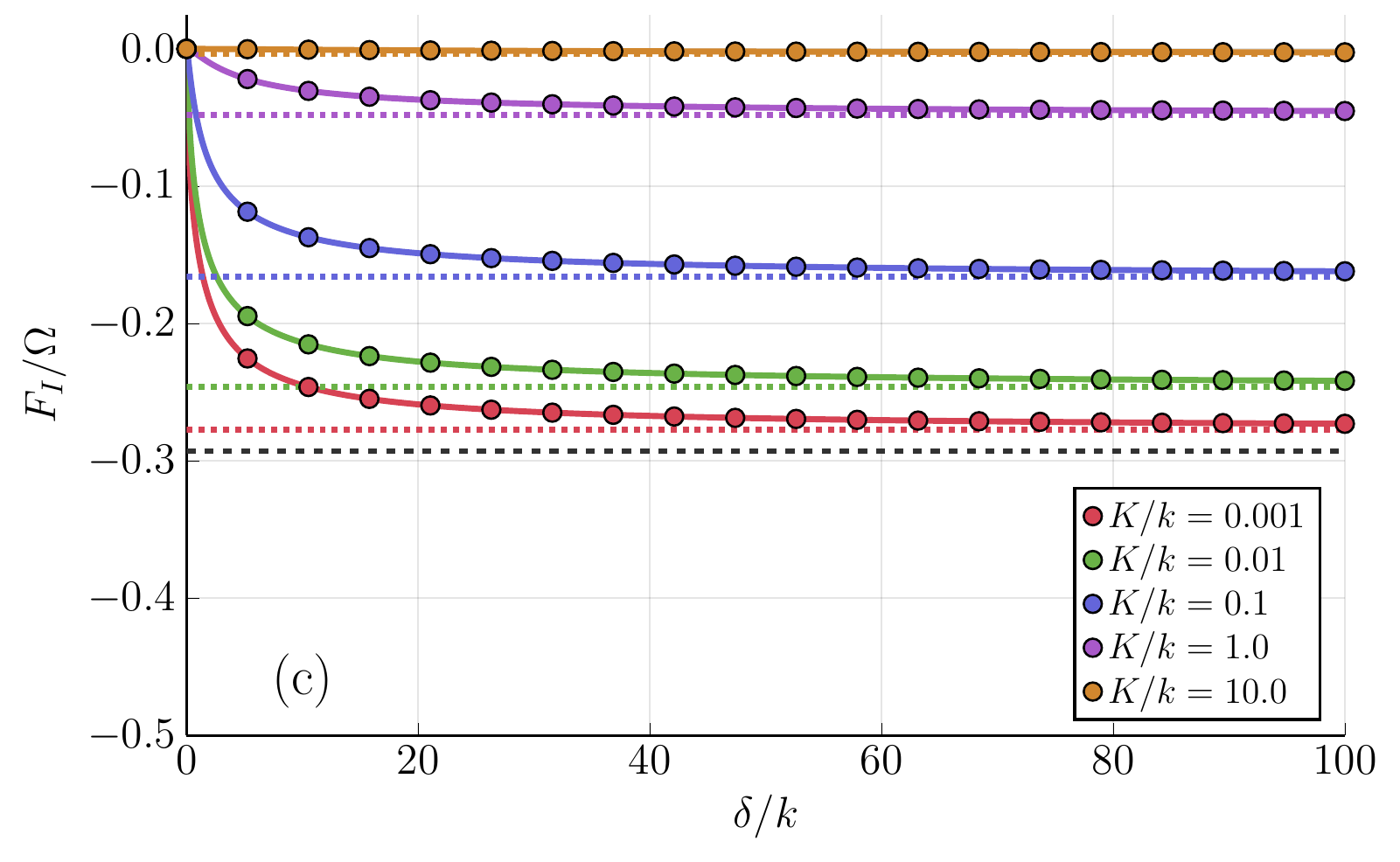}
    \caption{(a) $F_I$ for a diatomic molecule as a function of $M / m$ for $\delta = 0$ at $T = 0$. (b) Same as (a) but using a logarithmic scale on the $x$-axis. (c) $F_I$ as a function of $\delta / k$ for $M/ m = 1$ at $T = 0$. The markers correspond to the values obtained from our formalism, whereas the line plots are obtained from exact diagonalization. Unless stated otherwise, we follow this convention in the rest of the figures. The dotted lines mark the asymptotic values of $F_I$ as $M/m$ or $\delta /k$ approach infinity for the corresponding $K$. The dashed line is the asymptotic value for $K =0$. }
    \label{fig:Molecule_T0}
\end{figure}

The simplest system that one can study using our formalism is a diatomic molecule composed of two atoms of equal mass $m$ connected by a spring with the force constant $k$ and restricted to moving in one dimension. For the benefit of the subsequent discussion, we also confine each atom in an external harmonic potential $K$.

As there are only two atoms in the system, both will be subjected to perturbation, which we set to be the same for both. Thus, following the procedure given above, $\mathbf{m} = \mathrm{diag}(m\,m)$, $\Delta = \mathrm{diag}(\delta\,\delta)$, and $\Lambda = \mathrm{diag}(\lambda\,\lambda)$. Recall that $\lambda$ describes the change of the atomic mass $m\rightarrow M$ and is given by $M^{-1} - m^{-1}$.

Without the external perturbation, the mode frequencies and eigenvectors are obtained by solving
\begin{equation}
    \begin{pmatrix}
        k + K&-k
        \\
        -k &  k + K
    \end{pmatrix}\boldsymbol{\varepsilon}_s = \Omega_s^2 
    \begin{pmatrix}
        m&0
        \\
        0 & m
    \end{pmatrix}
    \boldsymbol{\varepsilon}_s\,.
    \label{eqn:diatomic_modes}
\end{equation}
This yields $\Omega_1^2 = K / m$ with $\boldsymbol{\varepsilon}_1 = (1,\, 1)/\sqrt{2}$ and $\Omega_2^2 =(2k+ K) / m$ with $\boldsymbol{\varepsilon}_2 = (1,\, -1)/\sqrt{2}$, leading to
\begin{align}
    \left[\Pi(z)\right]_{ij}&=
    \frac{ \boldsymbol{\varepsilon}_{1}^i  \boldsymbol{\varepsilon}_1^j }{-z^2+\Omega^2_{1}}
    +
    \frac{ \boldsymbol{\varepsilon}_{2}^i  \boldsymbol{\varepsilon}_{2}^j }{-z^2+\Omega^2_{2}}
    \nonumber
    \\
    &=
    \frac{1}{2}\left(\frac{1 }{-z^2+\Omega^2_{1}}
    \pm
    \frac{1 }{-z^2+\Omega^2_{2}}\right)\,,
    \label{eqn:Pi_molecule}
\end{align}
where $+$ corresponds to the diagonal elements of the $2\times 2$ $\Pi$ matrix and $-$ to the off-diagonal ones.

As the first step, we explore the system at zero temperature. Figure~\ref{fig:Molecule_T0}(a) shows $F_I$ as a function of $M / m$ for different values of $K$ for $\delta = 0$. To keep the relevant quantities dimensionless, we define $\Omega = \sqrt{k / m}$ as the characteristic energy scale. Solid lines lines are obtained from exact diagonalization while the symbols overlaying them correspond to the values computed using our formalism. For the $T = 0$ case, the summation over the Matsubara frequencies in Eq.~\eqref{eqn:F_I_Final} can be performed as a numerical integral.~\citep{Bruus2002} To calculate the frequencies using exact diagonalization, we replace zero, one, or two $m$'s in Eq.~\eqref{eqn:diatomic_modes} by $M$. We then compute the corresponding free energies using Eq.~\eqref{eqn:F_Bose}, from which we obtain $F_I$ by performing the subtraction described above. Figure~\ref{fig:Molecule_T0}(b) shows the same results as panel (a), but using a logarithmic scale for the $x$ axis to bring out the small-$M/m$ behavior. As expected, when $M = m$, $F_I = 0$ because this corresponds to a scenario where the atomic masses are unchanged. The interaction energy diverges as $M /m\rightarrow 0$ with small-$K$ systems exhibiting a faster divergence.

The dotted colored lines are the asymptotic values of $F_I$ as $M /m \rightarrow \infty$ and the dashed gray line is the asymptote for $K = 0$. These values are calculated using the mode frequencies for the two- and one-defect configurations in the $M / m\rightarrow \infty$ limit. In the former case, the frequencies of both modes go to zero, while in the latter one goes to zero and the other approaches $\Omega\sqrt{1+K/k}$. Given that for the unperturbed molecule, the frequencies are $\Omega\sqrt{K / k}$ and $\Omega\sqrt{2 + K/k}$, at $T = 0$, $F_I/\Omega \rightarrow (\sqrt{K / k}+\sqrt{2 + K/k}-2\sqrt{1 + K / k})/2$ as $M/m\rightarrow \infty$.

We also plot $F_I$ as a function of $\delta$ for $\lambda = 0$ in Fig.~\ref{fig:Molecule_T0}(c). Note that, for $\delta/k\rightarrow\infty$, the curves approach the same asymptotic values as in panel (a). This, of course, makes sense because extreme potential confinement $\delta/k\rightarrow \infty$ suppresses the motion of the atoms and is equivalent to replacing them with infinitely heavy immovable masses.

\begin{figure}
    \centering
    \includegraphics[width = \columnwidth]{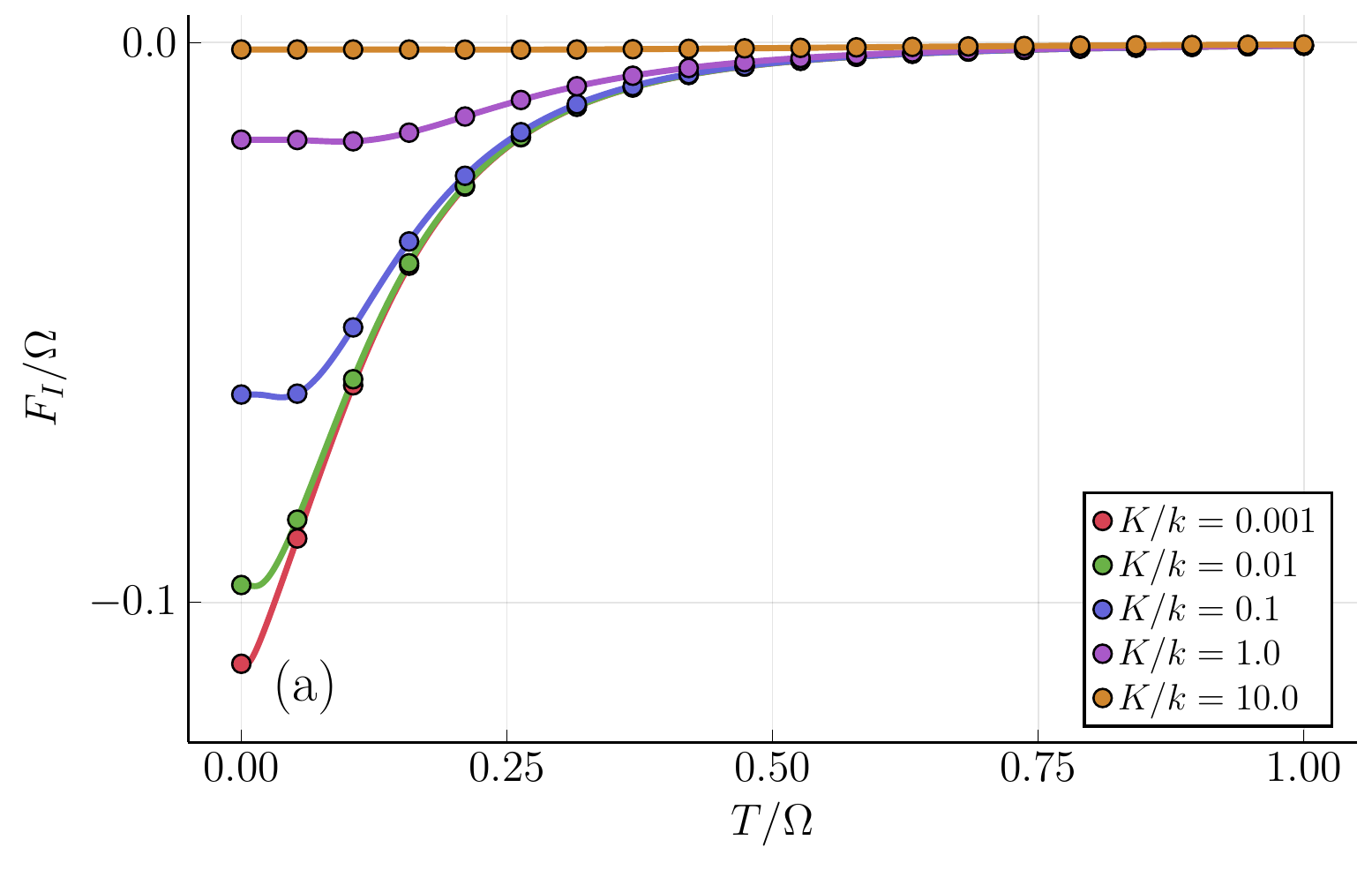}
    \includegraphics[width = \columnwidth]{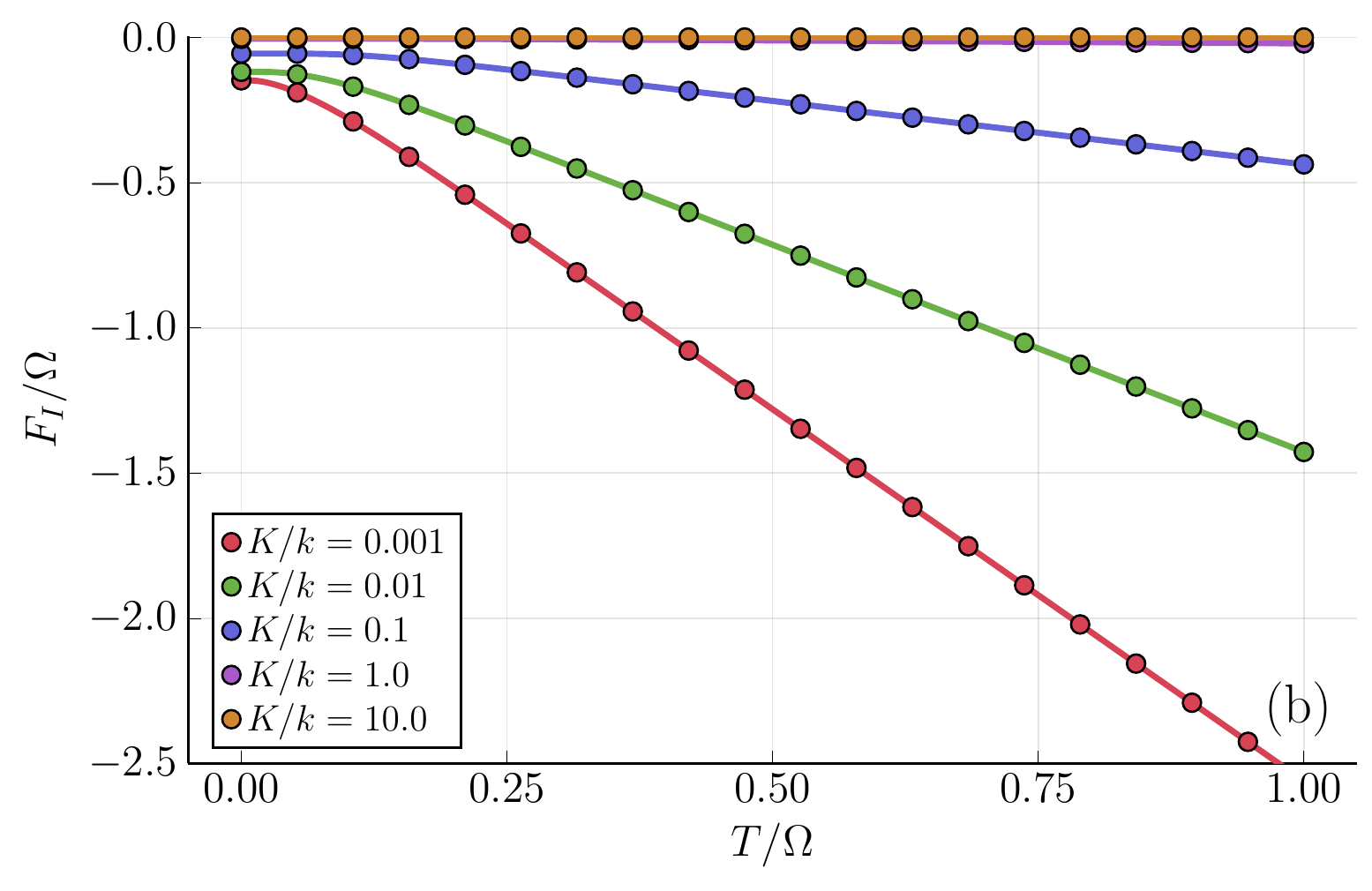}
    \includegraphics[width = \columnwidth] {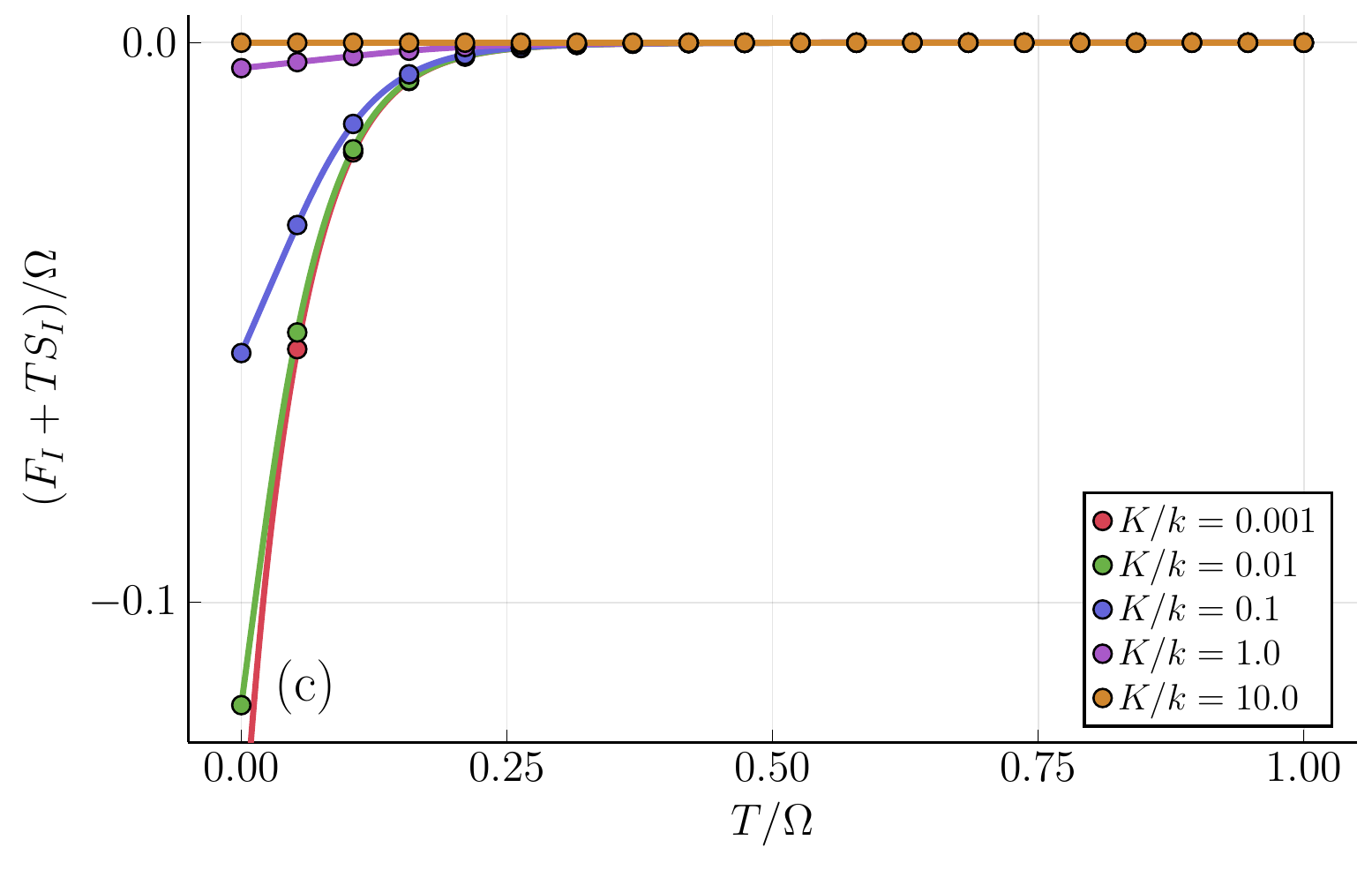}
    \caption{(a) $F_I$ for a diatomic molecule as a function of $T$ for $\delta = 0$ and $M/m = 10$. (b) $F_I$ as a function of $T$ for $M/ m = 1$ and $\delta/k = 1$. (c) $F_I + TS_I$ for the same system as panel (b).}
    \label{fig:Molecule}
\end{figure}

In addition to $T = 0$, we also compute $F_I$ at finite temperature. The exact diagonalization results are obtained from Eq.~\eqref{eqn:F_Bose} using the same steps as the $T = 0$ case. For the path integral approach, one can perform an integration along the real frequency axis.~\citep{Bruus2002} Alternatively, it is possible to sum a truncated series in Eq.~\eqref{eqn:F_I_Final}, which is the approach that we used by keeping the first 10,000 terms to guarantee a good agreement with the exact diagonalization results.

The confining potential $K$ plays an important role in numerical evaluation of Eq.~\eqref{eqn:F_Bose} at finite $T$. For $K = 0$, the system contains a zero-energy mode, leading to a divergence of the logarithm term. To mitigate this divergence, one can either drop the zero-mode or, as is done in this work, include a finite $K$.

Figure~\ref{fig:Molecule}(a) illustrates the decay of $F_I$ with increasing temperature for $M/m = 10$ and $\delta = 0$ at different $K$'s, in agreement with earlier studies.~\citep{Schecter2014pmc,Pavlov2018pmc, Pavlov2019, Rodin2019}. Panel (b) in Fig.~\ref{fig:Molecule} exhibits the linearly increasing $F_I$ for $\delta/k = 1$ and $M / m = 1$, as discussed in Sec.~\ref{sec:Exact_Diagonalization}, where we showed that this increase can be attributed to the entropy term in the free energy. By subtracting the zero-Matsubara-frequency term from $F_I$, we eliminate the linear behavior and obtain the expected decaying interaction, as seen in Fig.~\ref{fig:Molecule}(c).

\section{Periodic Systems}
\label{sec:Periodic_Systems}

To calculate $\Pi(z)$ used in computing the interaction energy, one needs to know the system's vibrational eigenstates and their corresponding eigenvalues, which involves diagonalizing the Hamiltonian. Hence, it might appear that the field theoretic approach offers no advantage over exact diagonalization and Eq.~\eqref{eqn:F_Bose}. In fact, it is worse because one needs to perform the Matsubara frequency summation. The true utility of our approach becomes clear when dealing with infinitely large systems.

Consider a $D$-dimensional Bravais lattice with $A$ atoms per unit cell and periodic boundary conditions, spanning $N$ unit cells along each of the basis vectors, where $N$ is assumed to be even. Diagonalizing such a system directly requires finding the eigenstates and eigenvalues of a $(ADN^D)\times(ADN^D)$ matrix, which is clearly not feasible as $N\rightarrow \infty$. This issue is especially pernicious in higher dimensionalities, where large-but-finite systems quickly become prohibitively expensive computationally. The field theoretic approach, on the other hand, can take advantage of the system periodicity to obtain the interaction in a straightforward manner.~\citep{Schecter2014pmc, Pavlov2018pmc, Pavlov2019, Rodin2019} Let us now demonstrate how our general result can be adapted to tackle periodic systems.

System periodicity requires that
\begin{equation}
    \boldsymbol{\varepsilon}_{s,j}\rightarrow
    \boldsymbol{\varepsilon}_{b,\{n\},a_j} 
    \prod_{l = 1}^D \sqrt{\frac{2}{N}}\mathrm{trig}_l\left(\frac{2\pi r_{j,l} n_l }{ N}\right)\,,
    \label{eqn:Periodic_Polarization}
\end{equation}
where $1\leq r_{j,l}\leq N$ is the integer coordinate of the unit cell hosting the $j$th atom along the $l$th basis vector. The polarization vector $\boldsymbol{\varepsilon}_{b, \{n\}}= (\boldsymbol{\varepsilon}_{b, \{n\}, 1}, ... \boldsymbol{\varepsilon}_{b, \{n\}, A})$ contains $AD$ elements and gives the relative motion of atoms within a unit cell for a particular mode. The subscript $b$ labels the phonon branch (of which there are $AD$), while $\{n\}$ is a set of harmonic numbers $0<n\leq N / 2$, collectively identifying a phonon mode. This polarization vector is computed by diagonalizing an $(AD)\times(AD)$ dynamical matrix for each set $\{n\}$,~\citep{Bruus2002} an obvious simplification compared to the $(ADN^D)\times(ADN^D)$ matrix for the exact diagonalization.

The amplitude of the oscillations for a given mode varies across the crystal in a periodic fashion, as dictated by the trigonometric function $\mathrm{trig}_l$, which can be either a sine or a cosine. One can see that the allowed values of $\{n\}$ provide the correct periodicity of these functions. Finally, the factor $\sqrt{2/N}$ guarantees that $\boldsymbol{\varepsilon}_s \cdot \boldsymbol{\varepsilon}_s= 1$. As a check, multiplying the number of combinations of the trigonometric functions $2^D$ by the number of harmonic indices $(N/2)^D$ and by the number of degrees of freedom $AD$ yields the correct number of modes $ADN^D$.

Using the fact that the mode frequency does not depend on the choice of the trigonometric function in Eq.~\eqref{eqn:Periodic_Polarization}, one gets
\begin{align}
  &\left[\Pi(z)\right]_{jk}=
  \sum_{b,\{n\}}\frac{ \boldsymbol{\varepsilon}_{b,\{n\},a_j} \otimes \boldsymbol{\varepsilon}_{b,\{n\},a_k} }{-z^2+\Omega^2_{b,\{n\}}}
  \nonumber
  \\
  \times &
  \prod_{l = 1}^D\frac{2}{N}
    \sum_{\{\mathrm{trig}_l\}}\left[
  \mathrm{trig}_l\left(\frac{2\pi r_{j,l} n_l }{ N}\right)
  \mathrm{trig}_l\left(\frac{2\pi r_{k,l} n_l }{ N}\right)
  \right]
  \nonumber
  \\
  =&
  \sum_{b,\{n\}}\frac{ \boldsymbol{\varepsilon}_{b,\{n\},a_j} \otimes \boldsymbol{\varepsilon}_{b,\{n\},a_k} }{-z^2+\Omega^2_{b,\{n\}}}
  \prod_{l = 1}^D\frac{2}{N}
  \cos\left(\frac{2\pi R_{jk,l} n_l }{ N}\right)
    \label{eqn:Pi_per}
\end{align}
with $ R_{jk,l} = r_{j,l} - r_{k,l}$. This form underscores the periodic nature of the system because only the separation between atoms $R_{jl,l}$ enters the expression, not their individual coordinates.

In the limit $N\rightarrow \infty$, we replace the summation over $\{n\}$ by integrals:
\begin{align}
  &\left[\Pi(z)\right]_{jk}=
  \sum_{b}\prod_{l=1}^D\oint d\theta_l\frac{ \boldsymbol{\varepsilon}_{b,\boldsymbol{\theta},a_j} \otimes \boldsymbol{\varepsilon}_{b,\boldsymbol{\theta},a_k} }{-z^2+\Omega^2_{b,\boldsymbol{\theta}}} \frac{\cos\left(R_{jk,l}\theta_l\right)}{2\pi}\,,
    \label{eqn:Pi_per_2}
\end{align}
where $\boldsymbol{\theta}$ is a vector of $\theta_l$. Because $\Omega^2_{b,\boldsymbol{\theta}}$ and $\boldsymbol{\varepsilon}_{b,\boldsymbol{\theta},a_j} \otimes \boldsymbol{\varepsilon}_{b,\boldsymbol{\theta},a_k}$ are even functions of $\boldsymbol{\theta}$, we can replace each of the cosines by exponentials to get a compact expression
\begin{align}
    \left[\Pi(z)\right]_{jk}=&\frac{1 }{(2\pi)^D} 
   \oint d\boldsymbol{\theta}
   e^{i\mathbf{R}_{jk}\cdot \boldsymbol{\theta}}
   \sum_{b}
   \frac{ \boldsymbol{\varepsilon}_{b,\boldsymbol{\theta},a_j} \otimes \boldsymbol{\varepsilon}_{b,\boldsymbol{\theta},a_k} }{-z^2+\Omega^2_{b,\boldsymbol{\theta}}}
  \,.
    \label{eqn:Pi_per_3}
\end{align}

One could have guessed the form of Eq.~\eqref{eqn:Pi_per_3} from Eq.~\eqref{eqn:Pi} by recalling that the eigenmodes in periodic systems are typically written as $\boldsymbol{\varepsilon}_{b,\mathbf{q}}e^{i\mathbf{r}\cdot\mathbf{q}}$. However, our derivation in Sec.~\ref{sec:General_Formalism} relied on the fact that $\boldsymbol{\varepsilon}_{s,l}$ were real, which is why we started with Eq.~\eqref{eqn:Periodic_Polarization} instead of just writing down Eq.~\eqref{eqn:Pi_per_3}.

\section{One-dimensional Chain}
\label{sec:1D_chain}

With the formalism for periodic systems established, we now demonstrate its application. To make the connection with prior work clear while highlighting the novelty provided by the new results, we apply it to a one-dimensional diatomic chain composed of alternating masses $m_1\mu$ and $m_2\mu$, where $\mu$ has units of mass and $m_{1/2}$ are dimensionless, connected by identical springs with the force constant $k$. As in the case of the diatomic molecule, the energy scale is set by $\Omega = \sqrt{k/\mu}$. The eigenmodes and their corresponding frequencies are obtained from
\begin{align}
   &\begin{pmatrix}
        m_1^{-\frac{1}{2}}&0
        \\
        0& m_2^{-\frac{1}{2}}
    \end{pmatrix}
    \begin{pmatrix}
        2 + K/k& - 1 - e^{-i\theta}
        \\
       - 1 - e^{i\theta} &2 + K/k
    \end{pmatrix}
    \nonumber
    \\
    \times&
    \begin{pmatrix}
        m_1^{-\frac{1}{2}}&0
        \\
        0& m_2^{-\frac{1}{2}}
    \end{pmatrix}
    \boldsymbol{\varepsilon}_\theta 
    =
    \frac{\Omega_\theta^2}{\Omega^2} \boldsymbol{\varepsilon}_\theta \,,
    \label{eqn:Chain_Equation}
\end{align}
where we included the confining harmonic potential $K$ like was done for the diatomic molecule.

Numerical diagonalization of Eq.~\eqref{eqn:Chain_Equation} yields the eigenmodes and the corresponding frequencies. Then, one picks out the required component of $\boldsymbol{\varepsilon}_\theta$ for each branch at a given $\theta$ and performs the branch summation, as shown in Eq.~\eqref{eqn:Pi_per_3}. Repeating the process for $\theta \in [0,2\pi]$ and taking the numerical integral over $\theta$ yields $\Pi(z)$. The remaining steps in calculating the interaction energy follow the procedure outlined in Sec.~\ref{sec:Free_Energy}. The summation over the Matsubara frequencies uses the approach of Sec.~\ref{sec:Diatomic_Molecule}: for $T = 0$, we integrate along the complex axis; for $T > 0$, we sum the first 10,000 terms in Eq.~\eqref{eqn:F_I_Final}.

An important advantage of using 1D systems to demonstrate the application of the new formalism is the possibility of validating the field theoretic results against exact diagonalization, for which we use periodic chains with $N = 1000$ unit cells. This length is sufficiently large to avoid the finite-size effects at the impurity separations considered here. To prevent the divergence of the finite-$T$ free energy, a small confining potential $K/k = 10^{-6}$ is included. As discussed above, to calculate the internal interaction energy between impurities, one drops the zero-frequency term from the Matsubara sum and subtracts the expression in Eq.~\eqref{eqn:F_I_Exact_Diagonalization} from $F_I$ calculated using exact diagonalization. For all the plots below, unless otherwise specified, the distances $d$ are given in terms of the interatomic spacing.  

\subsection{Monoatomic Chain}
\label{sec:monochain}

\begin{figure}
\includegraphics[width=\columnwidth]{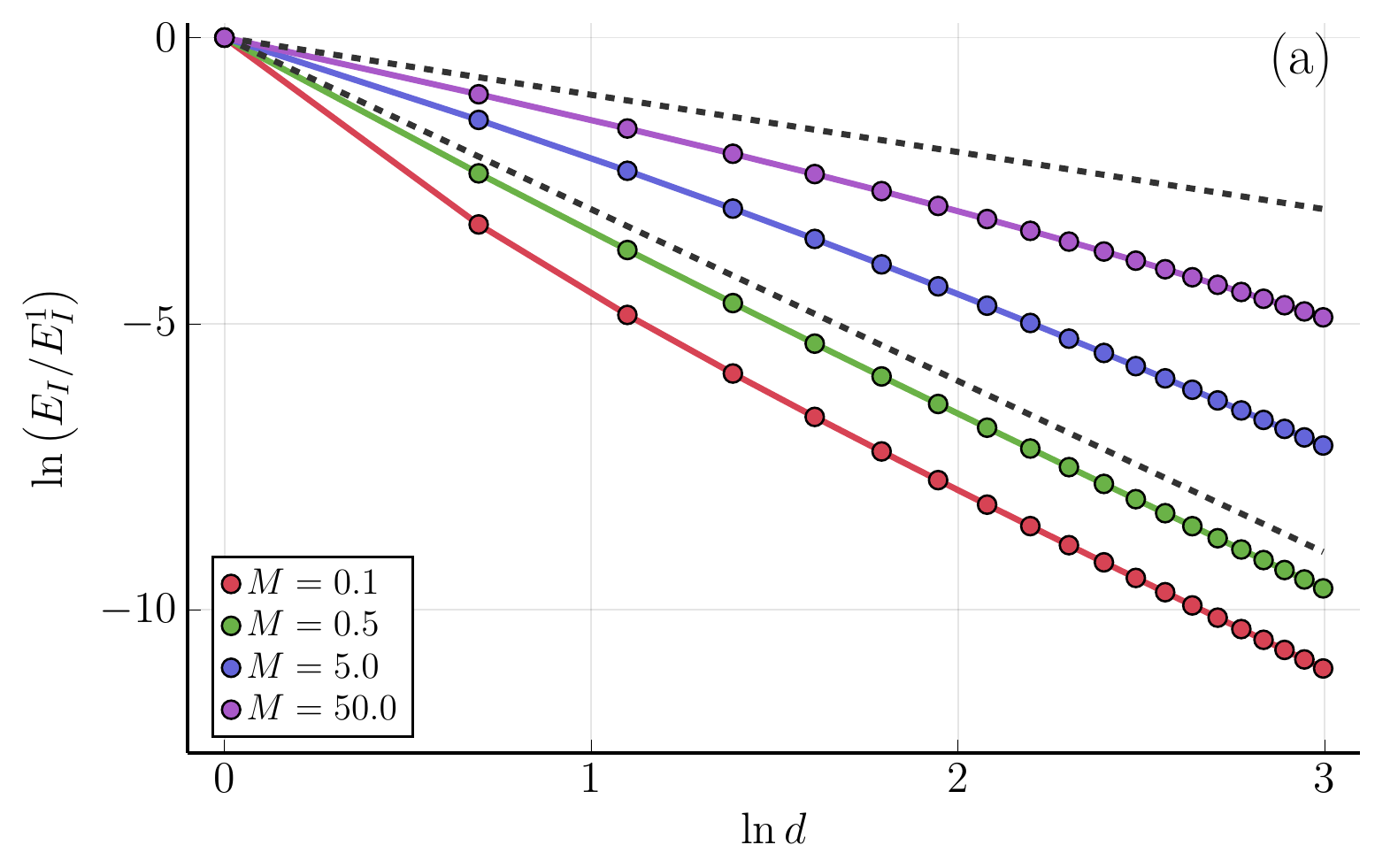}
\includegraphics[width=\columnwidth]{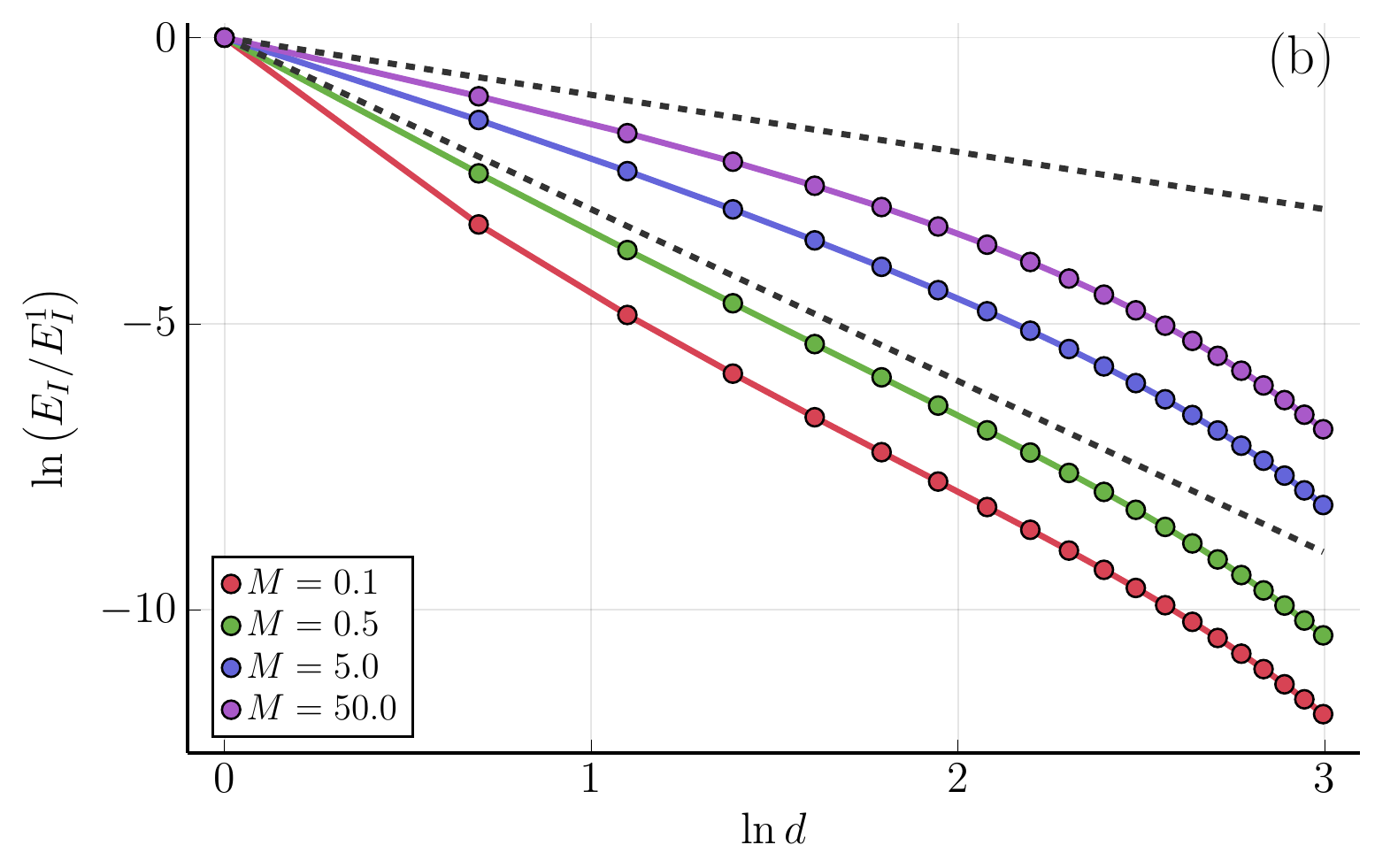}
\caption{(a) $\ln (E_I/E_I^1)$ at $T=0$ for two identical  impurities in a monoatomic chain, where $E_I^1$ is the interaction energy at separation $d = 1$.  (b) Same as panel $A$ at $T=0.02\Omega$. The masses of the impurities are indicated by the insets. The dashed lines follow $d^{-1}$ and $d^{-3}$ power laws.}
\label{fig:Monoatomic_M-M}
\end{figure}

We start by setting $m_1 = m_2 = 1$ to recover the monoatomic chain studied in earlier publications.~\citep{Schecter2014pmc, Pavlov2018pmc, Rodin2019} As the first example, we consider the interaction between pairs of identical impurities introduced by replacing two of the chain atoms by atoms with different masses. The interaction energy $E_I$ for several impurity masses $M\mu$ as a function of the defect separation $d$ is plotted in Fig.~\ref{fig:Monoatomic_M-M}, showing an excellent agreement between the path integral approach (markers) and exact diagonalization (lines). 

From Fig.~\ref{fig:Monoatomic_M-M} (a), one can see that the zero-temperature results are concordant with Refs.~\citep{Pavlov2018pmc, Rodin2019} demonstrating a quasi-power-law dependence of the interaction energy on the impurity separation. In the $d \gg 1$ limit, all the curves approach a slope of $-3$, in agreement with the inverse cubic interaction between mobile impurities at large distances.~\cite{Schecter2014pmc} At small $d$, increasing $M$ brings the slope closer to $-1$, expected in the case of stationary impurities. Conversely, if $M<1$, the interaction decays faster than the cube of the separation. Figure~\ref{fig:Monoatomic_M-M}(b) shows that increasing the temperature leads to a drastically faster decay of the interaction with distance, as expected.~\cite{Schecter2014pmc, Pavlov2019, Rodin2019}

\begin{figure}
\includegraphics[width=\columnwidth]{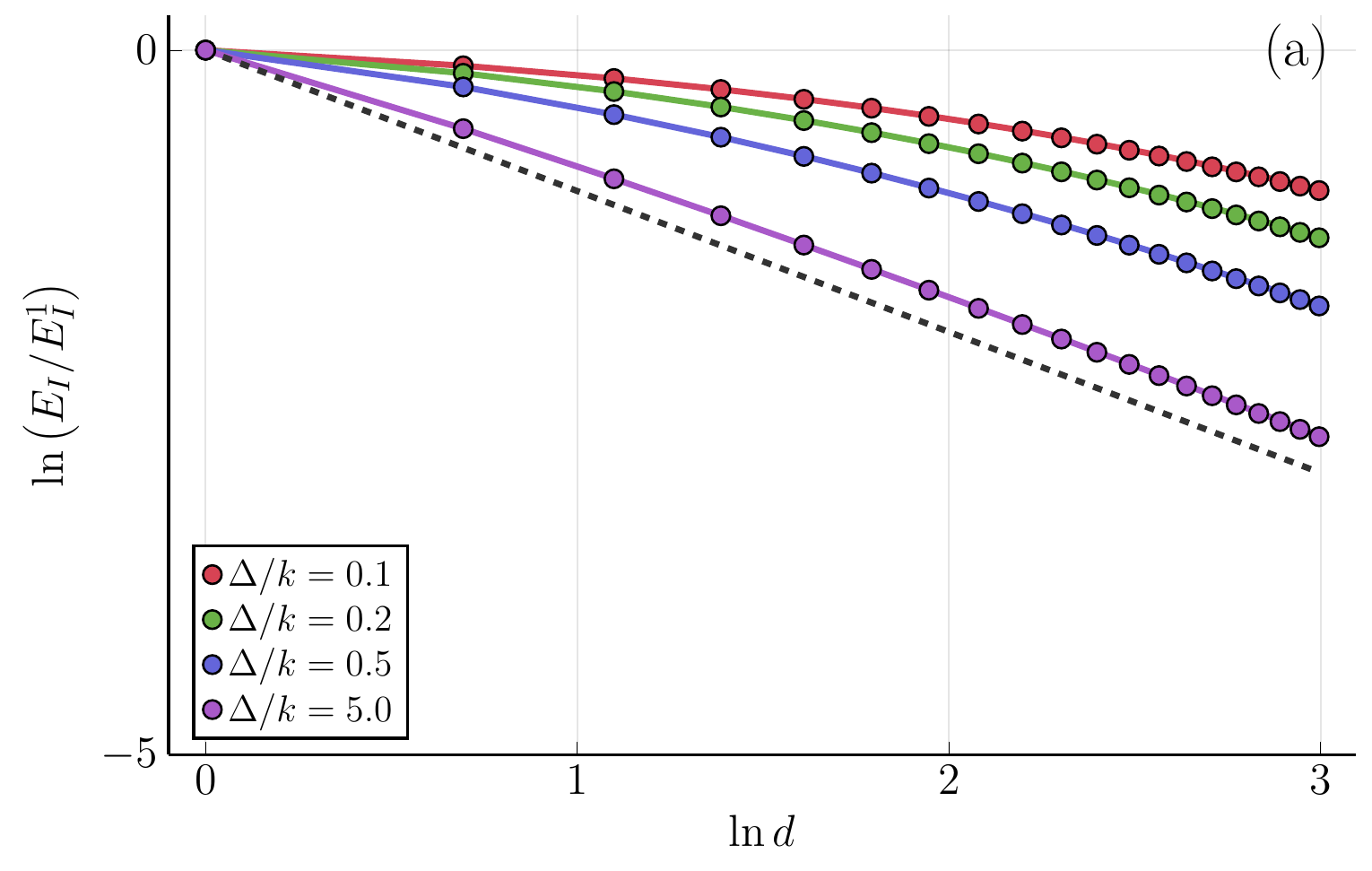}
\includegraphics[width=\columnwidth]{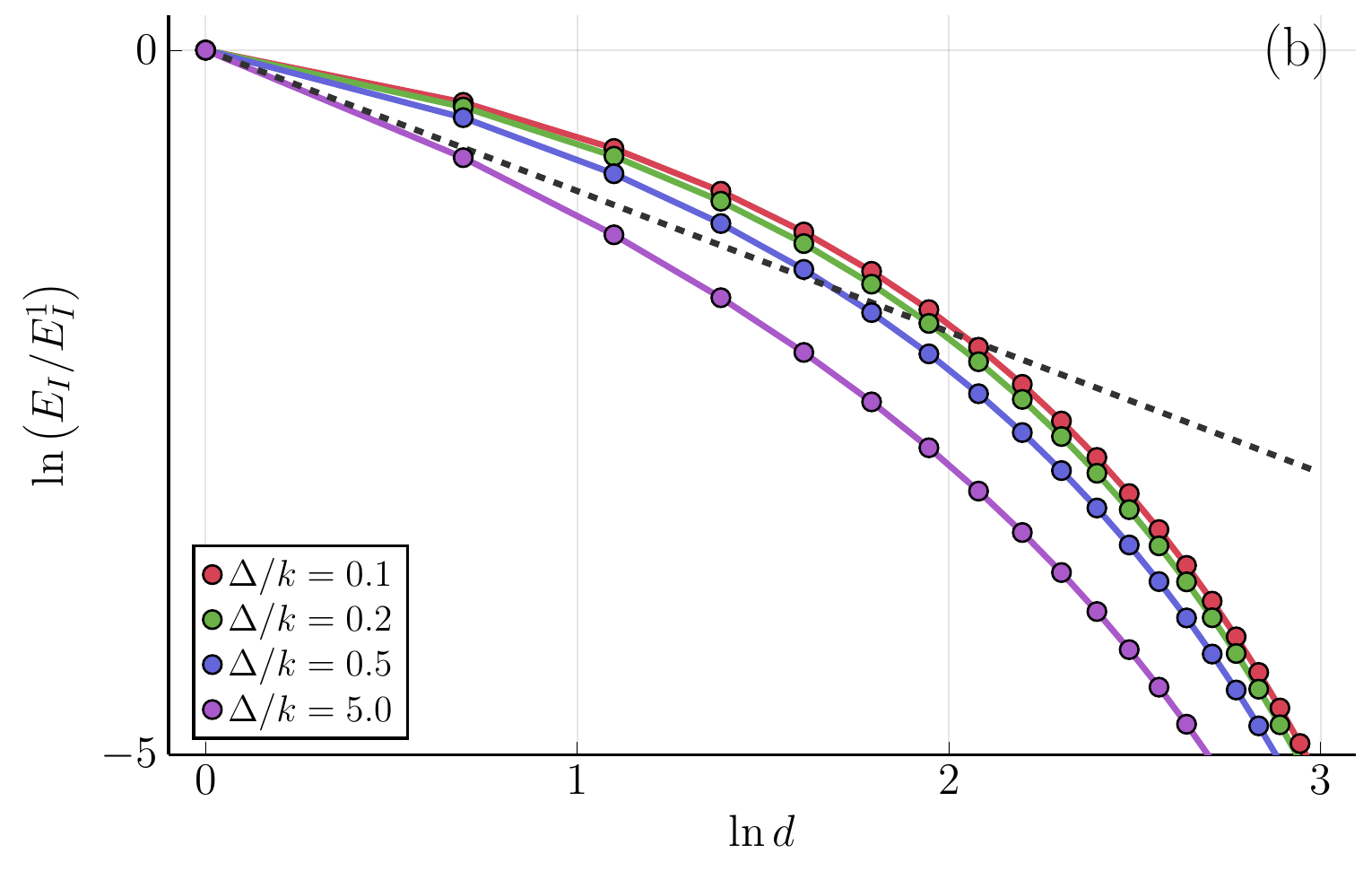}
\caption{(a) $\ln (E_I/E_I^1)$ at $T=0$ for two identical external potentials of various strengths in a monoatomic chain. (b) Same for $T = 0.02 \Omega$. The dashed lines are $d^{-1}$.}
\label{fig:Monoatomic_D-D}
\end{figure}

Following Refs.~\citep{Pavlov2018pmc, Pavlov2019}, we also investigate the interaction between chain atoms in external harmonic potentials. As for the impurity case, we check that our approach reproduces the previously known quasi-power-law dependence of energy on $d$ by plotting $E_I$ for two external potentials in Fig.~\ref{fig:Monoatomic_D-D}(a). The results show that for large values of $\Delta$, the interaction energy approaches $1 / d$ form from above and becomes slower as $\Delta$ is reduced, in agreement with Ref.~\citep{Pavlov2018pmc}.

In addition to addressing the zero-$T$ case, Fig.~\ref{fig:Monoatomic_D-D}(b) shows $E_I$ for two external potentials at finite $T$. Here, as before, the interaction loses its quasi-power law scaling with increased defect separation for all perturbation strengths. In contrast to the impurity pairs in Fig.~\ref{fig:Monoatomic_M-M}, the exponential suppression of the interaction by finite $T$ is evident at much smaller values of $d$. This behavior agrees with Ref.~\citep{Pavlov2019} as the $d^{-1}$-to-exponential transition for the external potentials is more drastic than the $d^{-3}$-to-exponential one for the impurities. Moreover, at large separations, $E_I$ for the potentials is suppressed by an additional $d^{-1}$ factor compared to the two-impurity case.

\begin{figure}
\includegraphics[width=\columnwidth]{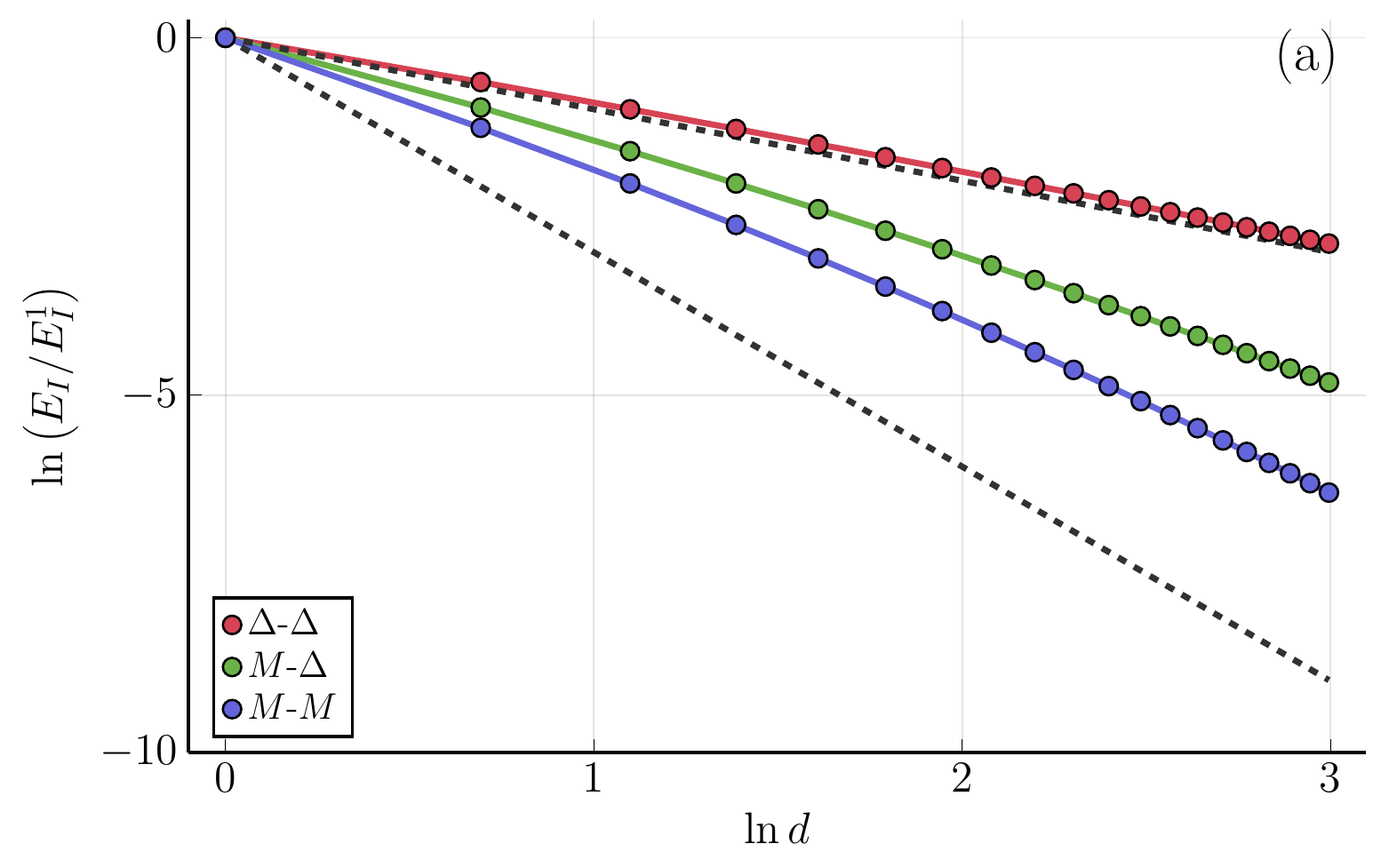}
\includegraphics[width=\columnwidth]{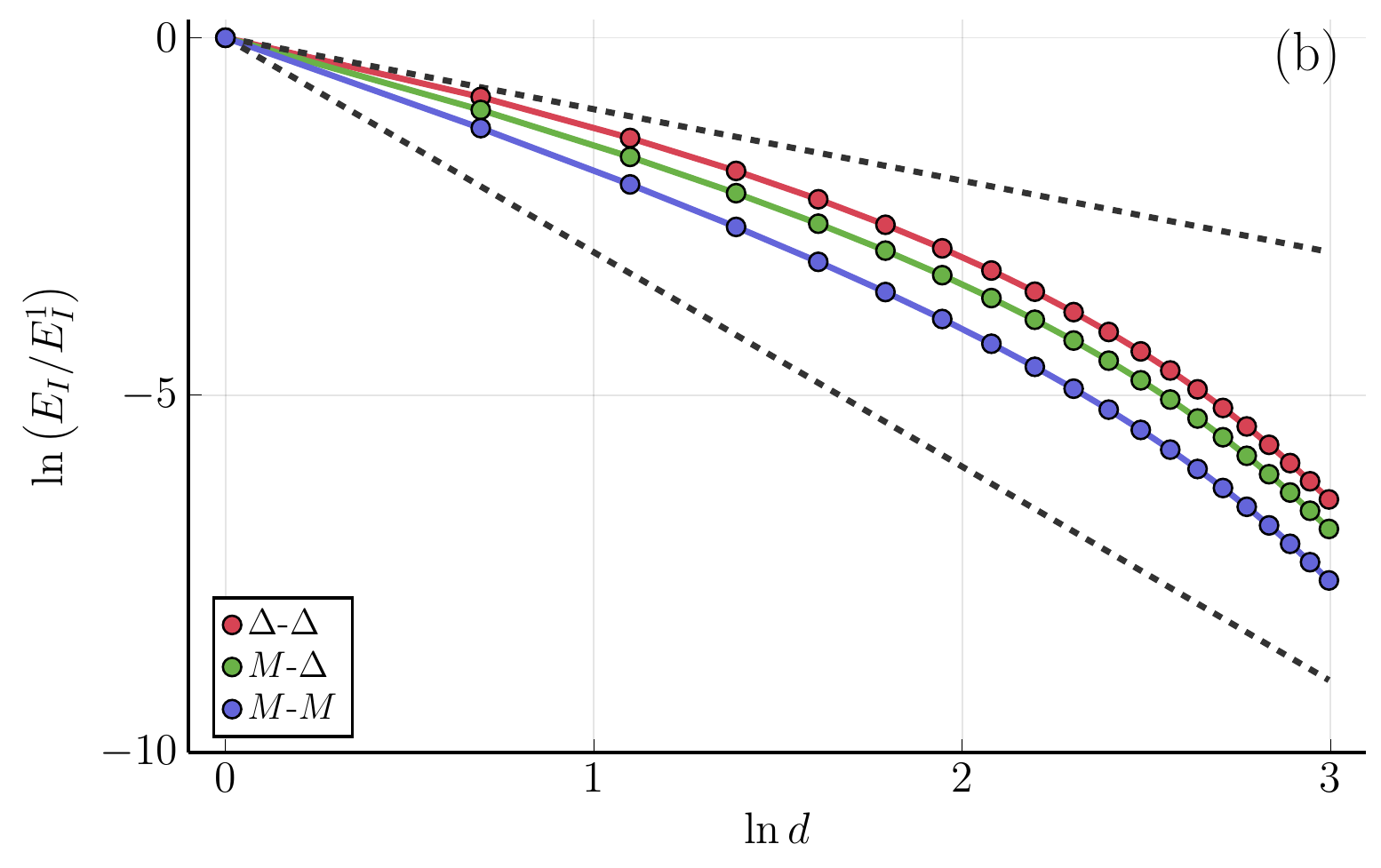}
\includegraphics[width = \columnwidth]{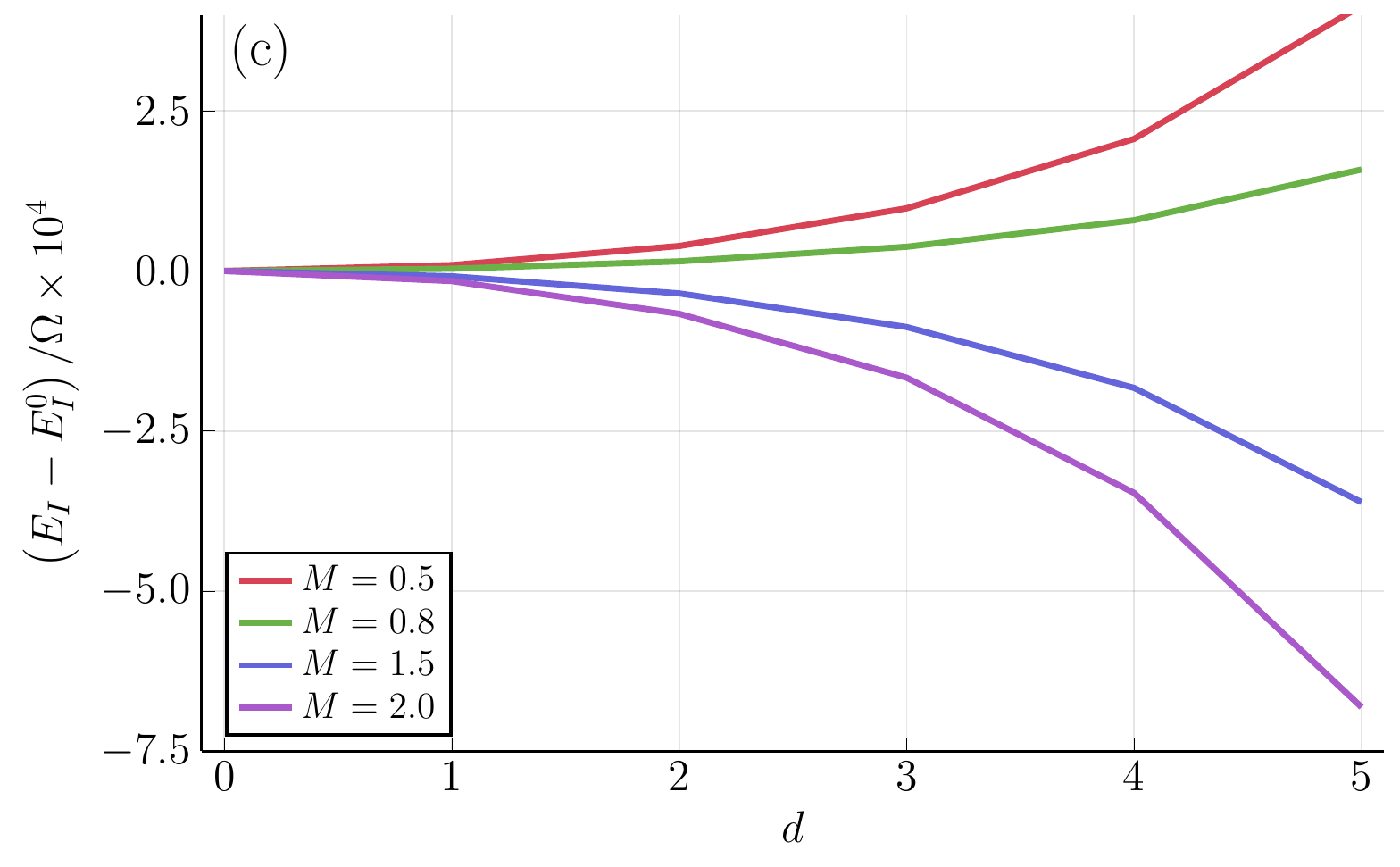}
\caption{(a) $\ln (E_I/E_I^1)$ at $T=0$ for two-impurity, two-potential, and mixed configurations with $M = \Delta / k = 10$. (b) Same for $T = 0.02 \Omega$. (c) $E_I - E_I^0$ vs. $d$ at $T = 0$ for a configuration where two external potentials are located on the $1$st and $19$th atoms of a chain with an impurity $M$ between them. $d$ is the distance from the midpoint ($10$th atom) and $E_I^0$ is the energy at $d = 0$. The results are obtained using the newly developed formalism.}
\label{fig:Mixed}
\end{figure}

Having demonstrated the ability of our formalism to reproduce known results, we use it to investigate the interaction between impurities and external potentials, which has not been addressed previously. To illustrate how $E_I$ for this ``mixed" configuration compares to the $E_I$'s for two impurities and two potentials, we plot the interaction energies for potential-impurity, two-impurity, and two-potential configurations with $M = \Delta / k = 10$ in Fig.~\ref{fig:Mixed}. As with earlier calculations, we validate the field theoretic results by comparing them to the ones obtained using exact diagonalization. For both zero [panel (a)] and finite [panel (b)] temperatures, $E_I / E_I^1$ for an impurity and an external potential lies strictly between the interaction energy for two impurities and two potentials. As expected, we observe that finite temperature induces a faster decay at higher values of $d$. 

It is also possible to investigate the behavior of clusters of impurities and external potentials. As an example, we calculate the energy profile of an impurity lying between two external potentials with $\Delta / k = 5$ located $19$ sites apart. Figure~\ref{fig:Mixed}(c) shows the dependence of the energy on the impurity's displacement from the midpoint between the two potentials. From the concavity of the curves, one observes that the midpoint is a stable equilibrium point if $M<1$ and unstable otherwise.

\begin{figure}
\includegraphics[width=\columnwidth]{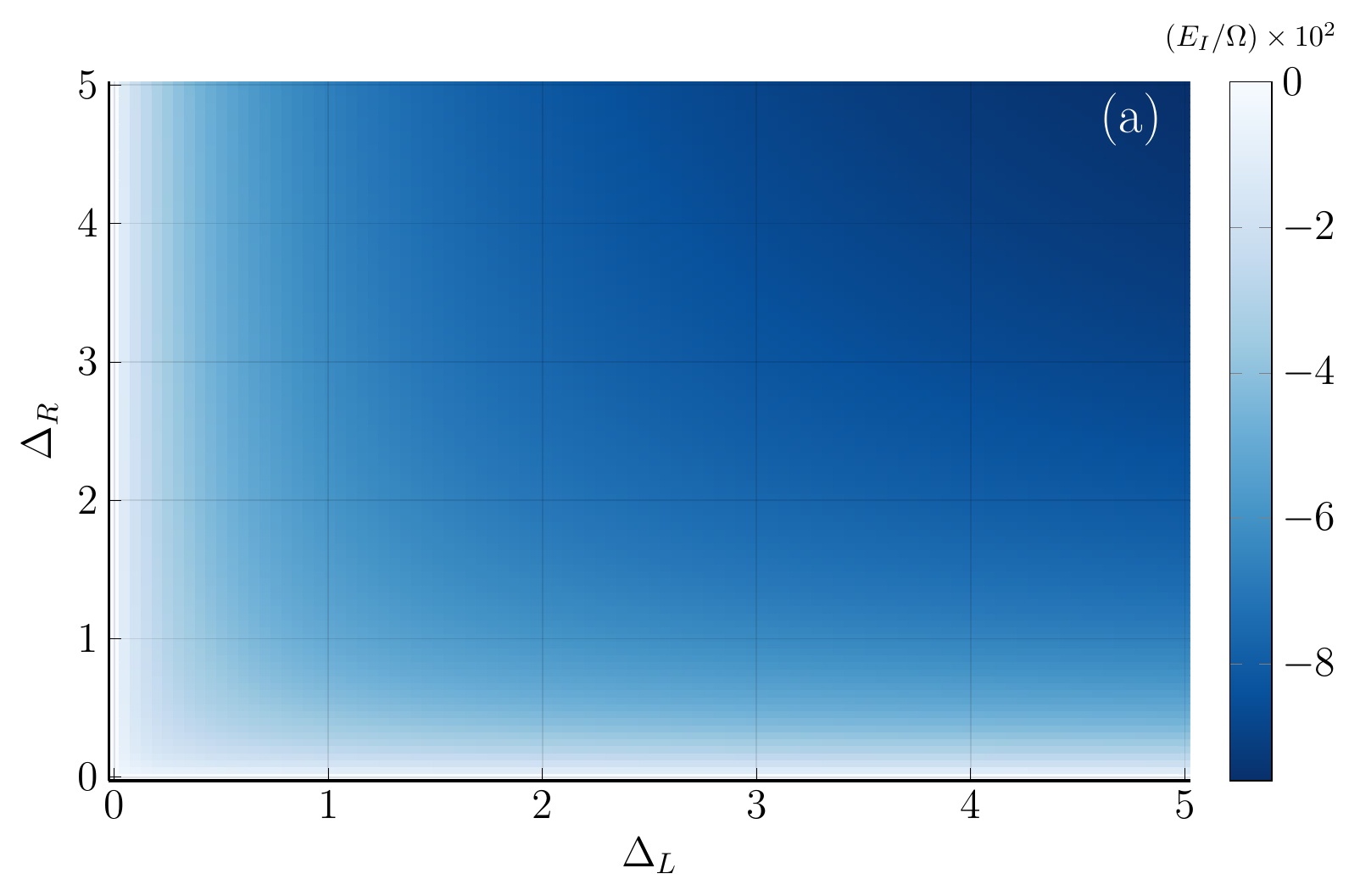}
\includegraphics[width=\columnwidth]{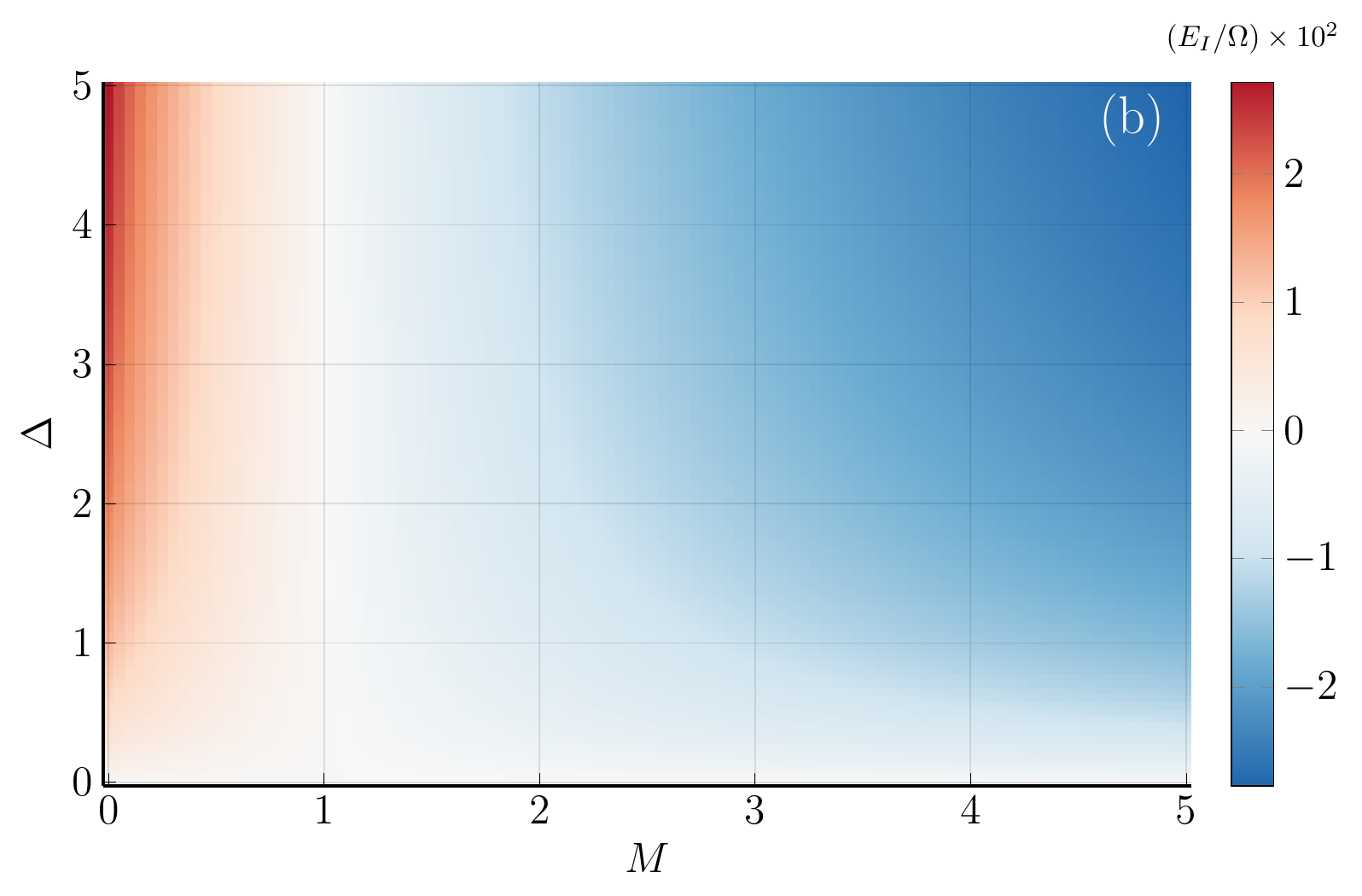}
\caption{Heat map of $E_I/\Omega$ at $T=0$ for (a) two externally-confined atoms and (b) for an impurity and an externally-confined atom in a monoatomic chain with separation $d=1$.}
\label{fig:Heatmap}
\end{figure}

Earlier work~\citep{Schecter2014pmc, Rodin2019} discussed the possibility of changing the sign of the PCE interaction for impurities by having one of them be lighter than the chain atom and the other one heavier. To extend this analysis to other defect combinations, we plot the interaction energy for pairs of adjacent defects in Fig.~\ref{fig:Heatmap}. Panel (a) shows that the interaction energy between two masses in external potentials is always negative. In contrast, Fig.~\ref{fig:Heatmap}(b) demonstrates that as if the impurity is lighter than the chain's atoms ($M<1$), its interaction with a mass in an external potential is repulsive, becoming attractive for $M>1$. This is consistent with Fig.~\ref{fig:Mixed}(c), where $M<1$ produces a stable equilibrium as the impurity is repelled by the externally confined atoms.

\subsection{Diatomic Chain}
\label{sec:dichain}

\begin{figure}
\includegraphics[width=\columnwidth]{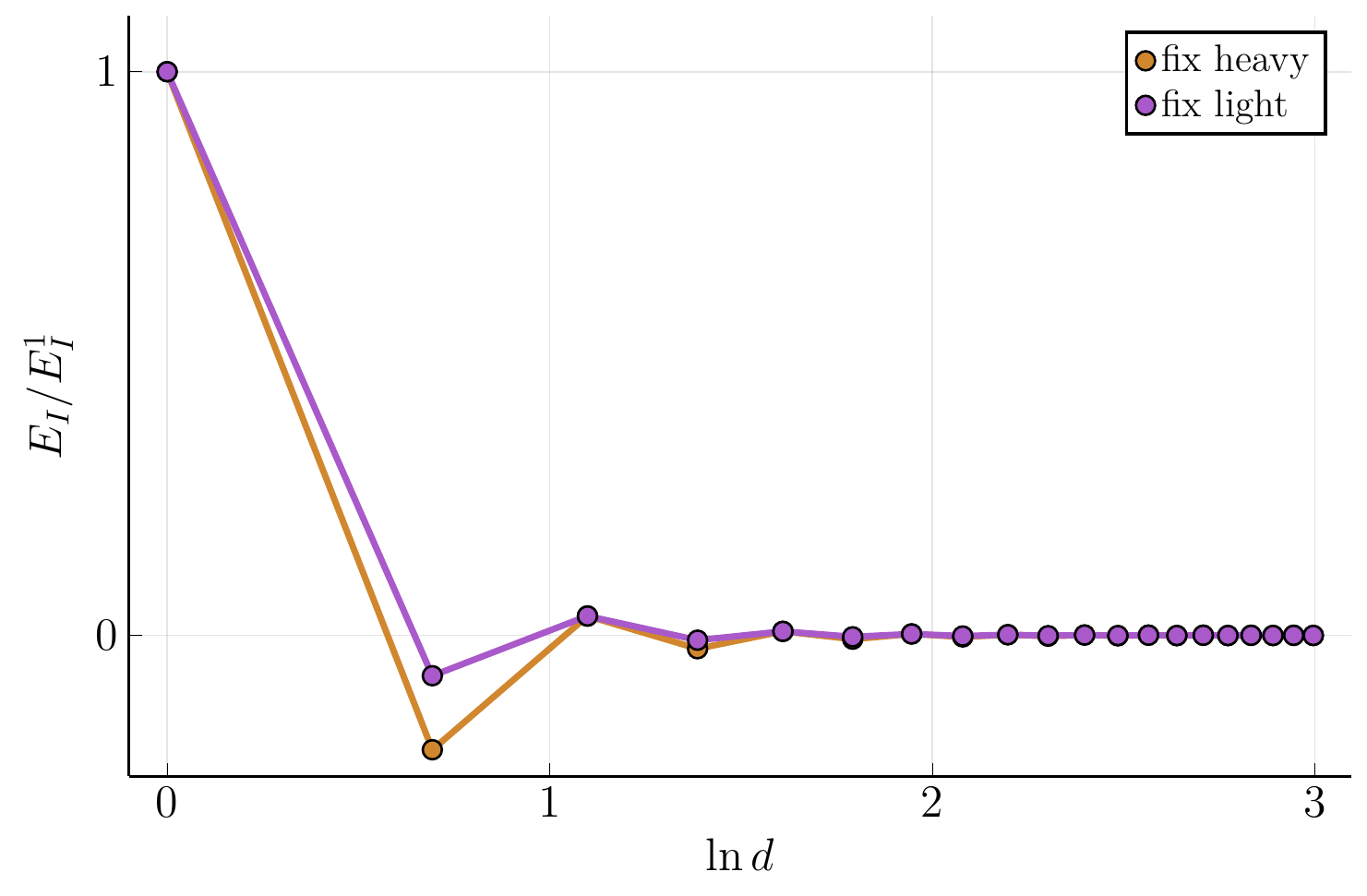}
\caption{$E_I/E_I^1$ at $T = 0$ for the interaction between two $M = 1.8$ impurities on a $m_1 = 1, m_2 = 3$ diatomic chain. "fix heavy (light)" means that one of the impurities is positioned at the $m_2$ ($m_1$) site and the second one is $d$ atoms away.}
\label{fig:Diatomic_Linear}
\end{figure}

\begin{figure}
\includegraphics[width=\columnwidth]{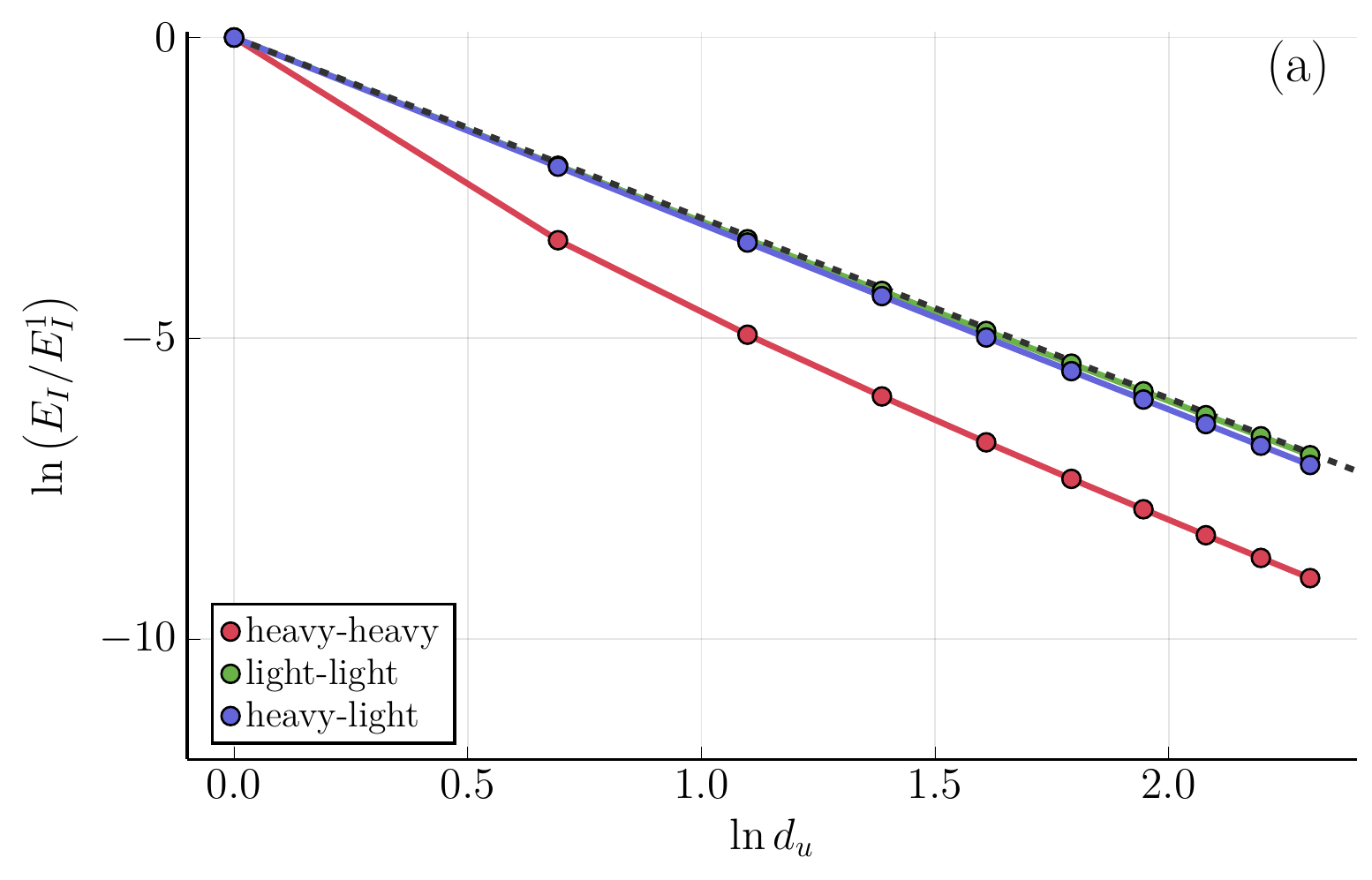}
\includegraphics[width=\columnwidth]{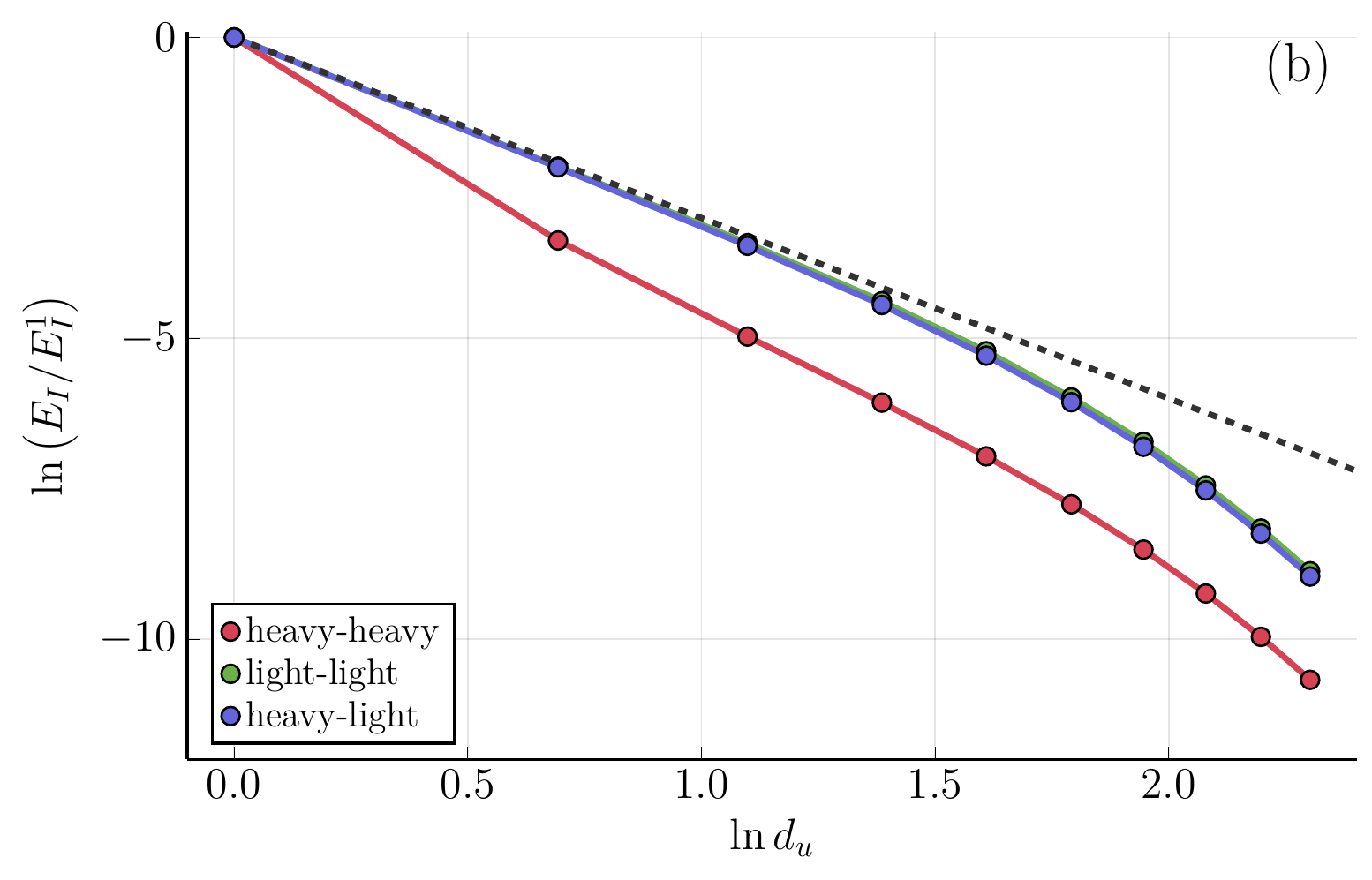}
\includegraphics[width = \columnwidth]{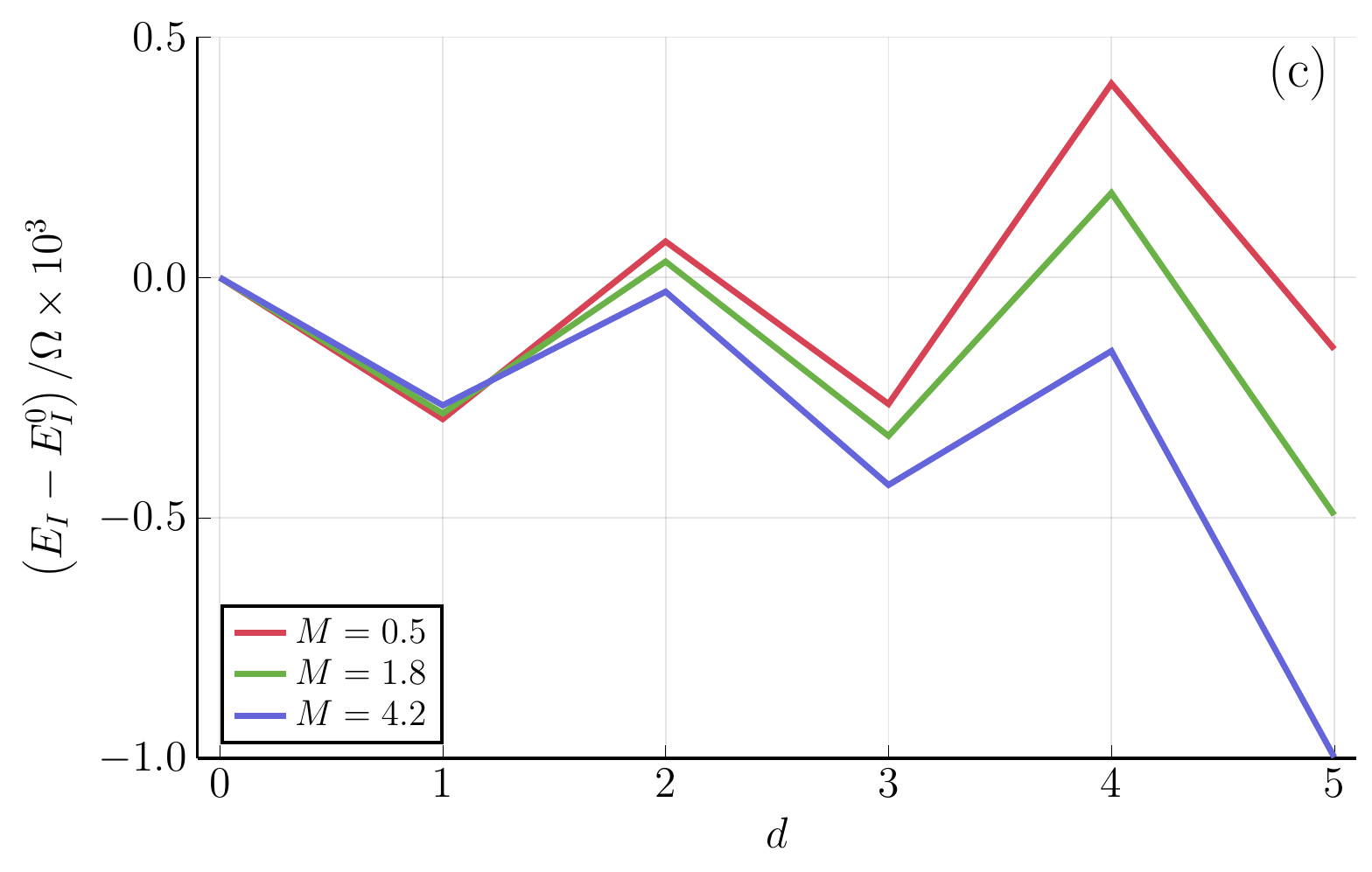}
\caption{(a) $\ln(E_I/E_I^1)$ at $T = 0$ for the interaction between two $M = 1.8$ impurities on a diatomic chain with $m_1 = 1$ and $m_2 = 3$. Here $d_u$ refers to the distance in terms of unit cells (hence twice the interatomic distance). (b) Same as (a) but for $T = 0.02 \Omega$. The dashed lines are $d_u^{-3}$. (c) $E_I - E_I^0$ vs. $d$ at $T = 0$ for a configuration where two external potentials are located on the $1$st and $19$th atoms of a chain with an impurity $M$ between them. $d$ is the distance from the midpoint ($10$th atom) and $E_I^0$ is the energy at $d = 0$. The central atom is a heavy atom, whilst the external potentials confine the light atoms. The results are obtained using the newly developed formalism.
}
\label{fig:Diatomic_Figs}
\end{figure}

For the final example, we consider a diatomic chain, consisting of alternating masses $m_1 = 1$ and $m_2 = 3$. As was mentioned earlier, in the monoatomic chain, it is known~\citep{Schecter2014pmc, Rodin2019} that the sign of the interaction between impurities is determined by whether they are both lighter or heavier than the chain's atoms. In a diatomic chain, we observe a more exotic version of this effect: for a pair of identical impurities with $m_1 < M < m_2$, we get an $E_I$ that changes sign with separation, as seen in Fig.~\ref{fig:Diatomic_Linear}. The interaction is always positive when one impurity replaces a heavy atom and the other replaces a light atom, and negative when they both replace the same kind of atom. It turns out that $E_I$ in each of these two regimes obey their own scaling laws,  resembling the $d^{-3}$ scaling of impurities in monoatomic chains, as seen in Fig.~\ref{fig:Diatomic_Figs}(a). Note that in contrast to the rest of the plots, for Fig.~\ref{fig:Diatomic_Figs} the unit of separation $d_u$ is measured in unit cells rather than interatomic separation. The scaling is thus in terms of $d_u$ rather than $d$.

Increasing the temperature does not alter the oscillatory form of $E_I$, nor the signs. At the same time the difference in the scaling from $T = 0$ is immediately clear from Fig.~\ref{fig:Diatomic_Figs}(b), where we plot the interaction energy at $T = 0.02\Omega$. Just as for the monoatomic chain, finite temperature leads to an accelerated decay of $E_I$ as compared to the $T = 0$ case. 

Finally, we investigate clusters formed by one impurity between two external potentials on the diatomic chain. The energy profile of this setup is plotted in Fig.~\ref{fig:Diatomic_Figs}(c). Compared to the interaction energy of clusters in the monoatomic chain, the energy landscape shown here is more uneven and the diatomic structure of the underlying system can be observed. Since the central atom in this case is a heavy atom, odd $d$s correspond to the impurity residing at the site of a light atom. The oscillating nature of $E_I$ turns these sites into local traps.

\section{Conclusions}
\label{sec:Conclusions}

In summary, we have employed the path integral formalism to derive an expression for the finite-temperature Helmholtz free energy in a general system with vibrational modes in the presence of defects. Specifically, this approach can handle impurities, external potentials, or their combinations. Our results make it possible to extract the non-pairwise interaction energy between defects. We have also shown how one can compute the internal interaction energy, as well as entropy using our approach.

As a demonstration of our method, we performed a series of calculations on a diatomic molecule, as well as mono- and diatomic chains. We validated our results by comparing them against exact diagonalization calculations and earlier known results. It is important to stress that while exact diagonalization can be faster than our approach for finite-$T$ calculations in one-dimensional chains, modeling ``infinite" systems becomes prohibitively expensive in higher dimensions (1000 unit cells are needed for a hardly-infinite $10\times10\times10$ cube). Therefore, the approach developed in this paper is especially useful in higher-dimensional scenarios with large defect separation which increases the minimum system size for exact diagonalization.

\section*{Acknowledgments}

The numerical calculations were performed using JULIA programming language~\citep{Bezanson2017}. The code is available at \href{https://github.com/rodin-physics/phonon-casimir-1d-optical}{https://github.com/rodin-physics/phonon-casimir-1d-optical}. The authors express their gratitude to Keian Noori for his help with the manuscript preparation. The authors acknowledge the National Research Foundation, Prime Minister Office, Singapore, under its Medium Sized Centre Programme and the support by Yale-NUS College (through Grant No. R-607-265-380-121).

\bibliography{Phonon_Casimir.bib}

\begin{thebibliography}{28}%
\makeatletter
\providecommand \@ifxundefined [1]{%
 \@ifx{#1\undefined}
}%
\providecommand \@ifnum [1]{%
 \ifnum #1\expandafter \@firstoftwo
 \else \expandafter \@secondoftwo
 \fi
}%
\providecommand \@ifx [1]{%
 \ifx #1\expandafter \@firstoftwo
 \else \expandafter \@secondoftwo
 \fi
}%
\providecommand \natexlab [1]{#1}%
\providecommand \enquote  [1]{``#1''}%
\providecommand \bibnamefont  [1]{#1}%
\providecommand \bibfnamefont [1]{#1}%
\providecommand \citenamefont [1]{#1}%
\providecommand \href@noop [0]{\@secondoftwo}%
\providecommand \href [0]{\begingroup \@sanitize@url \@href}%
\providecommand \@href[1]{\@@startlink{#1}\@@href}%
\providecommand \@@href[1]{\endgroup#1\@@endlink}%
\providecommand \@sanitize@url [0]{\catcode `\\12\catcode `\$12\catcode
  `\&12\catcode `\#12\catcode `\^12\catcode `\_12\catcode `\%12\relax}%
\providecommand \@@startlink[1]{}%
\providecommand \@@endlink[0]{}%
\providecommand \url  [0]{\begingroup\@sanitize@url \@url }%
\providecommand \@url [1]{\endgroup\@href {#1}{\urlprefix }}%
\providecommand \urlprefix  [0]{URL }%
\providecommand \Eprint [0]{\href }%
\providecommand \doibase [0]{https://doi.org/}%
\providecommand \selectlanguage [0]{\@gobble}%
\providecommand \bibinfo  [0]{\@secondoftwo}%
\providecommand \bibfield  [0]{\@secondoftwo}%
\providecommand \translation [1]{[#1]}%
\providecommand \BibitemOpen [0]{}%
\providecommand \bibitemStop [0]{}%
\providecommand \bibitemNoStop [0]{.\EOS\space}%
\providecommand \EOS [0]{\spacefactor3000\relax}%
\providecommand \BibitemShut  [1]{\csname bibitem#1\endcsname}%
\let\auto@bib@innerbib\@empty
\bibitem [{\citenamefont {Casimir}(1948)}]{Casimir1948ota}%
  \BibitemOpen
  \bibfield  {author} {\bibinfo {author} {\bibfnamefont {H.~B.~G.}\
  \bibnamefont {Casimir}},\ }\href
  {http://www.dwc.knaw.nl/DL/publications/PU00018547.pdf} {\bibfield  {journal}
  {\bibinfo  {journal} {Proc. K. Ned. Akad. B-Ph.}\ }\textbf {\bibinfo {volume}
  {51}},\ \bibinfo {pages} {793} (\bibinfo {year} {1948})}\BibitemShut
  {NoStop}%
\bibitem [{\citenamefont {Lamoreaux}(1997)}]{Lamoreaux1997}%
  \BibitemOpen
  \bibfield  {author} {\bibinfo {author} {\bibfnamefont {S.~K.}\ \bibnamefont
  {Lamoreaux}},\ }\href {https://doi.org/10.1103/PhysRevLett.78.5} {\bibfield
  {journal} {\bibinfo  {journal} {Physical Review Letters}\ }\textbf {\bibinfo
  {volume} {78}},\ \bibinfo {pages} {5} (\bibinfo {year} {1997})}\BibitemShut
  {NoStop}%
\bibitem [{\citenamefont {Bressi}\ \emph {et~al.}(2002)\citenamefont {Bressi},
  \citenamefont {Carugno}, \citenamefont {Onofrio}, \citenamefont {Ruoso},
  \citenamefont {Onofrio},\ and\ \citenamefont {Ruoso}}]{Bressi2002}%
  \BibitemOpen
  \bibfield  {author} {\bibinfo {author} {\bibfnamefont {G.}~\bibnamefont
  {Bressi}}, \bibinfo {author} {\bibfnamefont {G.}~\bibnamefont {Carugno}},
  \bibinfo {author} {\bibfnamefont {R.}~\bibnamefont {Onofrio}}, \bibinfo
  {author} {\bibfnamefont {G.}~\bibnamefont {Ruoso}}, \bibinfo {author}
  {\bibfnamefont {R.}~\bibnamefont {Onofrio}},\ and\ \bibinfo {author}
  {\bibfnamefont {G.}~\bibnamefont {Ruoso}},\ }\href
  {https://doi.org/10.1103/PhysRevLett.88.041804} {\bibfield  {journal}
  {\bibinfo  {journal} {Physical Review Letters}\ }\textbf {\bibinfo {volume}
  {88}},\ \bibinfo {pages} {041804} (\bibinfo {year} {2002})},\ \Eprint
  {https://arxiv.org/abs/0203002} {arXiv:0203002 [quant-ph]} \BibitemShut
  {NoStop}%
\bibitem [{\citenamefont {Krause}\ \emph {et~al.}(2007)\citenamefont {Krause},
  \citenamefont {Decca}, \citenamefont {L{\'{o}}pez},\ and\ \citenamefont
  {Fischbach}}]{Krause2007}%
  \BibitemOpen
  \bibfield  {author} {\bibinfo {author} {\bibfnamefont {D.~E.}\ \bibnamefont
  {Krause}}, \bibinfo {author} {\bibfnamefont {R.~S.}\ \bibnamefont {Decca}},
  \bibinfo {author} {\bibfnamefont {D.}~\bibnamefont {L{\'{o}}pez}},\ and\
  \bibinfo {author} {\bibfnamefont {E.}~\bibnamefont {Fischbach}},\ }\href
  {https://doi.org/10.1103/PhysRevLett.98.050403} {\bibfield  {journal}
  {\bibinfo  {journal} {Phys. Rev. Lett.}\ }\textbf {\bibinfo {volume} {98}},\
  \bibinfo {pages} {050403} (\bibinfo {year} {2007})}\BibitemShut {NoStop}%
\bibitem [{\citenamefont {Klimchitskaya}\ \emph {et~al.}(2009)\citenamefont
  {Klimchitskaya}, \citenamefont {Mohideen},\ and\ \citenamefont
  {Mostepanenko}}]{Klimchitskaya2009}%
  \BibitemOpen
  \bibfield  {author} {\bibinfo {author} {\bibfnamefont {G.~L.}\ \bibnamefont
  {Klimchitskaya}}, \bibinfo {author} {\bibfnamefont {U.}~\bibnamefont
  {Mohideen}},\ and\ \bibinfo {author} {\bibfnamefont {V.~M.}\ \bibnamefont
  {Mostepanenko}},\ }\href {https://doi.org/10.1103/RevModPhys.81.1827}
  {\bibfield  {journal} {\bibinfo  {journal} {Rev. Mod. Phys}\ }\textbf
  {\bibinfo {volume} {81}},\ \bibinfo {pages} {1827} (\bibinfo {year}
  {2009})},\ \Eprint {https://arxiv.org/abs/0902.4022} {arXiv:0902.4022}
  \BibitemShut {NoStop}%
\bibitem [{\citenamefont {Munday}\ \emph {et~al.}(2009)\citenamefont {Munday},
  \citenamefont {Capasso},\ and\ \citenamefont {Parsegian}}]{Munday2009}%
  \BibitemOpen
  \bibfield  {author} {\bibinfo {author} {\bibfnamefont {J.~N.}\ \bibnamefont
  {Munday}}, \bibinfo {author} {\bibfnamefont {F.}~\bibnamefont {Capasso}},\
  and\ \bibinfo {author} {\bibfnamefont {V.~A.}\ \bibnamefont {Parsegian}},\
  }\href {https://doi.org/10.1038/nature07610} {\bibfield  {journal} {\bibinfo
  {journal} {Nature}\ }\textbf {\bibinfo {volume} {457}},\ \bibinfo {pages}
  {170} (\bibinfo {year} {2009})}\BibitemShut {NoStop}%
\bibitem [{\citenamefont {French}\ \emph {et~al.}(2010)\citenamefont {French},
  \citenamefont {Parsegian}, \citenamefont {Podgornik}, \citenamefont {Rajter},
  \citenamefont {Jagota}, \citenamefont {Luo}, \citenamefont {Asthagiri},
  \citenamefont {Chaudhury}, \citenamefont {Chiang}, \citenamefont {Granick},
  \citenamefont {Kalinin}, \citenamefont {Kardar}, \citenamefont {Kjellander},
  \citenamefont {Langreth}, \citenamefont {Lewis}, \citenamefont {Lustig},
  \citenamefont {Wesolowski}, \citenamefont {Wettlaufer}, \citenamefont
  {Ching}, \citenamefont {Finnis}, \citenamefont {Houlihan}, \citenamefont
  {{Von Lilienfeld}}, \citenamefont {{Van Oss}},\ and\ \citenamefont
  {Zemb}}]{French2010}%
  \BibitemOpen
  \bibfield  {author} {\bibinfo {author} {\bibfnamefont {R.~H.}\ \bibnamefont
  {French}}, \bibinfo {author} {\bibfnamefont {V.~A.}\ \bibnamefont
  {Parsegian}}, \bibinfo {author} {\bibfnamefont {R.}~\bibnamefont
  {Podgornik}}, \bibinfo {author} {\bibfnamefont {R.~F.}\ \bibnamefont
  {Rajter}}, \bibinfo {author} {\bibfnamefont {A.}~\bibnamefont {Jagota}},
  \bibinfo {author} {\bibfnamefont {J.}~\bibnamefont {Luo}}, \bibinfo {author}
  {\bibfnamefont {D.}~\bibnamefont {Asthagiri}}, \bibinfo {author}
  {\bibfnamefont {M.~K.}\ \bibnamefont {Chaudhury}}, \bibinfo {author}
  {\bibfnamefont {Y.~M.}\ \bibnamefont {Chiang}}, \bibinfo {author}
  {\bibfnamefont {S.}~\bibnamefont {Granick}}, \bibinfo {author} {\bibfnamefont
  {S.}~\bibnamefont {Kalinin}}, \bibinfo {author} {\bibfnamefont
  {M.}~\bibnamefont {Kardar}}, \bibinfo {author} {\bibfnamefont
  {R.}~\bibnamefont {Kjellander}}, \bibinfo {author} {\bibfnamefont {D.~C.}\
  \bibnamefont {Langreth}}, \bibinfo {author} {\bibfnamefont {J.}~\bibnamefont
  {Lewis}}, \bibinfo {author} {\bibfnamefont {S.}~\bibnamefont {Lustig}},
  \bibinfo {author} {\bibfnamefont {D.}~\bibnamefont {Wesolowski}}, \bibinfo
  {author} {\bibfnamefont {J.~S.}\ \bibnamefont {Wettlaufer}}, \bibinfo
  {author} {\bibfnamefont {W.~Y.}\ \bibnamefont {Ching}}, \bibinfo {author}
  {\bibfnamefont {M.}~\bibnamefont {Finnis}}, \bibinfo {author} {\bibfnamefont
  {F.}~\bibnamefont {Houlihan}}, \bibinfo {author} {\bibfnamefont {O.~A.}\
  \bibnamefont {{Von Lilienfeld}}}, \bibinfo {author} {\bibfnamefont {C.~J.}\
  \bibnamefont {{Van Oss}}},\ and\ \bibinfo {author} {\bibfnamefont
  {T.}~\bibnamefont {Zemb}},\ }\href
  {https://doi.org/10.1103/RevModPhys.82.1887} {\bibfield  {journal} {\bibinfo
  {journal} {Rev. Mod. Phys}\ }\textbf {\bibinfo {volume} {82}},\ \bibinfo
  {pages} {1887} (\bibinfo {year} {2010})}\BibitemShut {NoStop}%
\bibitem [{\citenamefont {Sushkov}\ \emph {et~al.}(2011)\citenamefont
  {Sushkov}, \citenamefont {Kim}, \citenamefont {Dalvit},\ and\ \citenamefont
  {Lamoreaux}}]{Sushkov2011}%
  \BibitemOpen
  \bibfield  {author} {\bibinfo {author} {\bibfnamefont {A.~O.}\ \bibnamefont
  {Sushkov}}, \bibinfo {author} {\bibfnamefont {W.~J.}\ \bibnamefont {Kim}},
  \bibinfo {author} {\bibfnamefont {D.~A.}\ \bibnamefont {Dalvit}},\ and\
  \bibinfo {author} {\bibfnamefont {S.~K.}\ \bibnamefont {Lamoreaux}},\ }\href
  {https://doi.org/10.1038/nphys1909} {\bibfield  {journal} {\bibinfo
  {journal} {Nature Physics}\ }\textbf {\bibinfo {volume} {7}},\ \bibinfo
  {pages} {230} (\bibinfo {year} {2011})},\ \Eprint
  {https://arxiv.org/abs/1011.5219} {arXiv:1011.5219} \BibitemShut {NoStop}%
\bibitem [{\citenamefont {Rodriguez}\ \emph {et~al.}(2011)\citenamefont
  {Rodriguez}, \citenamefont {Capasso},\ and\ \citenamefont
  {Johnson}}]{Rodriguez2011}%
  \BibitemOpen
  \bibfield  {author} {\bibinfo {author} {\bibfnamefont {A.~W.}\ \bibnamefont
  {Rodriguez}}, \bibinfo {author} {\bibfnamefont {F.}~\bibnamefont {Capasso}},\
  and\ \bibinfo {author} {\bibfnamefont {S.~G.}\ \bibnamefont {Johnson}},\
  }\href {https://doi.org/10.1038/nphoton.2011.39} {\bibfield  {journal}
  {\bibinfo  {journal} {Nature Photonics}\ }\textbf {\bibinfo {volume} {5}},\
  \bibinfo {pages} {211} (\bibinfo {year} {2011})}\BibitemShut {NoStop}%
\bibitem [{\citenamefont {Zou}\ \emph {et~al.}(2013)\citenamefont {Zou},
  \citenamefont {Marcet}, \citenamefont {Rodriguez}, \citenamefont {Reid},
  \citenamefont {McCauley}, \citenamefont {Kravchenko}, \citenamefont {Lu},
  \citenamefont {Bao}, \citenamefont {Johnson},\ and\ \citenamefont
  {Chan}}]{Zou2013}%
  \BibitemOpen
  \bibfield  {author} {\bibinfo {author} {\bibfnamefont {J.}~\bibnamefont
  {Zou}}, \bibinfo {author} {\bibfnamefont {Z.}~\bibnamefont {Marcet}},
  \bibinfo {author} {\bibfnamefont {A.~W.}\ \bibnamefont {Rodriguez}}, \bibinfo
  {author} {\bibfnamefont {M.~T.}\ \bibnamefont {Reid}}, \bibinfo {author}
  {\bibfnamefont {A.~P.}\ \bibnamefont {McCauley}}, \bibinfo {author}
  {\bibfnamefont {I.~I.}\ \bibnamefont {Kravchenko}}, \bibinfo {author}
  {\bibfnamefont {T.}~\bibnamefont {Lu}}, \bibinfo {author} {\bibfnamefont
  {Y.}~\bibnamefont {Bao}}, \bibinfo {author} {\bibfnamefont {S.~G.}\
  \bibnamefont {Johnson}},\ and\ \bibinfo {author} {\bibfnamefont {H.~B.}\
  \bibnamefont {Chan}},\ }\href {https://doi.org/10.1038/ncomms2842} {\bibfield
   {journal} {\bibinfo  {journal} {Nature Communications}\ }\textbf {\bibinfo
  {volume} {4}},\ \bibinfo {pages} {1} (\bibinfo {year} {2013})},\ \Eprint
  {https://arxiv.org/abs/1207.6163} {arXiv:1207.6163} \BibitemShut {NoStop}%
\bibitem [{\citenamefont {Intravaia}\ \emph {et~al.}(2013)\citenamefont
  {Intravaia}, \citenamefont {Koev}, \citenamefont {Jung}, \citenamefont
  {Talin}, \citenamefont {Davids}, \citenamefont {Decca}, \citenamefont
  {Aksyuk}, \citenamefont {Dalvit},\ and\ \citenamefont
  {L{\'{o}}pez}}]{Intravaia2013}%
  \BibitemOpen
  \bibfield  {author} {\bibinfo {author} {\bibfnamefont {F.}~\bibnamefont
  {Intravaia}}, \bibinfo {author} {\bibfnamefont {S.}~\bibnamefont {Koev}},
  \bibinfo {author} {\bibfnamefont {I.~W.}\ \bibnamefont {Jung}}, \bibinfo
  {author} {\bibfnamefont {A.~A.}\ \bibnamefont {Talin}}, \bibinfo {author}
  {\bibfnamefont {P.~S.}\ \bibnamefont {Davids}}, \bibinfo {author}
  {\bibfnamefont {R.~S.}\ \bibnamefont {Decca}}, \bibinfo {author}
  {\bibfnamefont {V.~A.}\ \bibnamefont {Aksyuk}}, \bibinfo {author}
  {\bibfnamefont {D.~A.}\ \bibnamefont {Dalvit}},\ and\ \bibinfo {author}
  {\bibfnamefont {D.}~\bibnamefont {L{\'{o}}pez}},\ }\href
  {https://doi.org/10.1038/ncomms3515} {\bibfield  {journal} {\bibinfo
  {journal} {Nat. Commun.}\ }\textbf {\bibinfo {volume} {4}},\ \bibinfo {pages}
  {1} (\bibinfo {year} {2013})},\ \Eprint {https://arxiv.org/abs/1202.6356}
  {arXiv:1202.6356} \BibitemShut {NoStop}%
\bibitem [{\citenamefont {Garrett}\ \emph {et~al.}(2019)\citenamefont
  {Garrett}, \citenamefont {Somers}, \citenamefont {Sendgikoski},\ and\
  \citenamefont {Munday}}]{Garrett2019}%
  \BibitemOpen
  \bibfield  {author} {\bibinfo {author} {\bibfnamefont {J.~L.}\ \bibnamefont
  {Garrett}}, \bibinfo {author} {\bibfnamefont {D.~A.}\ \bibnamefont {Somers}},
  \bibinfo {author} {\bibfnamefont {K.}~\bibnamefont {Sendgikoski}},\ and\
  \bibinfo {author} {\bibfnamefont {J.~N.}\ \bibnamefont {Munday}},\ }\href
  {https://doi.org/10.1103/PhysRevA.100.022508} {\bibfield  {journal} {\bibinfo
   {journal} {Phys. Rev. A}\ }\textbf {\bibinfo {volume} {100}},\ \bibinfo
  {pages} {22508} (\bibinfo {year} {2019})},\ \Eprint
  {https://arxiv.org/abs/1811.07175} {arXiv:1811.07175} \BibitemShut {NoStop}%
\bibitem [{\citenamefont {Fong}\ \emph {et~al.}(2019)\citenamefont {Fong},
  \citenamefont {Li}, \citenamefont {Zhao}, \citenamefont {Yang}, \citenamefont
  {Wang},\ and\ \citenamefont {Zhang}}]{Fong2019}%
  \BibitemOpen
  \bibfield  {author} {\bibinfo {author} {\bibfnamefont {K.~Y.}\ \bibnamefont
  {Fong}}, \bibinfo {author} {\bibfnamefont {H.~K.}\ \bibnamefont {Li}},
  \bibinfo {author} {\bibfnamefont {R.}~\bibnamefont {Zhao}}, \bibinfo {author}
  {\bibfnamefont {S.}~\bibnamefont {Yang}}, \bibinfo {author} {\bibfnamefont
  {Y.}~\bibnamefont {Wang}},\ and\ \bibinfo {author} {\bibfnamefont
  {X.}~\bibnamefont {Zhang}},\ }\href
  {https://doi.org/10.1038/s41586-019-1800-4} {\bibfield  {journal} {\bibinfo
  {journal} {Nature}\ }\textbf {\bibinfo {volume} {576}},\ \bibinfo {pages}
  {243} (\bibinfo {year} {2019})}\BibitemShut {NoStop}%
\bibitem [{\citenamefont {Moritz}\ \emph {et~al.}(2003)\citenamefont {Moritz},
  \citenamefont {St{\"{o}}ferle}, \citenamefont {K{\"{o}}hl},\ and\
  \citenamefont {Esslinger}}]{Moritz2003}%
  \BibitemOpen
  \bibfield  {author} {\bibinfo {author} {\bibfnamefont {H.}~\bibnamefont
  {Moritz}}, \bibinfo {author} {\bibfnamefont {T.}~\bibnamefont
  {St{\"{o}}ferle}}, \bibinfo {author} {\bibfnamefont {M.}~\bibnamefont
  {K{\"{o}}hl}},\ and\ \bibinfo {author} {\bibfnamefont {T.}~\bibnamefont
  {Esslinger}},\ }\href {https://doi.org/10.1103/PhysRevLett.91.250402}
  {\bibfield  {journal} {\bibinfo  {journal} {Phys. Rev. Lett.}\ }\textbf
  {\bibinfo {volume} {91}},\ \bibinfo {pages} {250402} (\bibinfo {year}
  {2003})},\ \Eprint {https://arxiv.org/abs/0307607} {arXiv:0307607 [cond-mat]}
  \BibitemShut {NoStop}%
\bibitem [{\citenamefont {Tolra}\ \emph {et~al.}(2004)\citenamefont {Tolra},
  \citenamefont {O'Hara}, \citenamefont {Huckans}, \citenamefont {Phillips},
  \citenamefont {Rolston},\ and\ \citenamefont {Porto}}]{Tolra2004}%
  \BibitemOpen
  \bibfield  {author} {\bibinfo {author} {\bibfnamefont {B.~L.}\ \bibnamefont
  {Tolra}}, \bibinfo {author} {\bibfnamefont {K.~M.}\ \bibnamefont {O'Hara}},
  \bibinfo {author} {\bibfnamefont {J.~H.}\ \bibnamefont {Huckans}}, \bibinfo
  {author} {\bibfnamefont {W.~D.}\ \bibnamefont {Phillips}}, \bibinfo {author}
  {\bibfnamefont {S.~L.}\ \bibnamefont {Rolston}},\ and\ \bibinfo {author}
  {\bibfnamefont {J.~V.}\ \bibnamefont {Porto}},\ }\href
  {https://doi.org/10.1103/PhysRevLett.92.190401} {\bibfield  {journal}
  {\bibinfo  {journal} {Phys. Rev. Lett.}\ }\textbf {\bibinfo {volume} {92}},\
  \bibinfo {pages} {190401} (\bibinfo {year} {2004})}\BibitemShut {NoStop}%
\bibitem [{\citenamefont {Moritz}\ \emph {et~al.}(2005)\citenamefont {Moritz},
  \citenamefont {St{\"{o}}ferle}, \citenamefont {G{\"{u}}nter}, \citenamefont
  {K{\"{o}}hl},\ and\ \citenamefont {Esslinger}}]{Moritz2005}%
  \BibitemOpen
  \bibfield  {author} {\bibinfo {author} {\bibfnamefont {H.}~\bibnamefont
  {Moritz}}, \bibinfo {author} {\bibfnamefont {T.}~\bibnamefont
  {St{\"{o}}ferle}}, \bibinfo {author} {\bibfnamefont {K.}~\bibnamefont
  {G{\"{u}}nter}}, \bibinfo {author} {\bibfnamefont {M.}~\bibnamefont
  {K{\"{o}}hl}},\ and\ \bibinfo {author} {\bibfnamefont {T.}~\bibnamefont
  {Esslinger}},\ }\href {https://doi.org/10.1103/PhysRevLett.94.210401}
  {\bibfield  {journal} {\bibinfo  {journal} {Phys. Rev. Lett.}\ }\textbf
  {\bibinfo {volume} {94}},\ \bibinfo {pages} {210401} (\bibinfo {year}
  {2005})},\ \Eprint {https://arxiv.org/abs/0503202} {arXiv:0503202 [cond-mat]}
  \BibitemShut {NoStop}%
\bibitem [{\citenamefont {Catani}\ \emph {et~al.}(2012)\citenamefont {Catani},
  \citenamefont {Lamporesi}, \citenamefont {Naik}, \citenamefont {Gring},
  \citenamefont {Inguscio}, \citenamefont {Minardi}, \citenamefont {Kantian},\
  and\ \citenamefont {Giamarchi}}]{Catani2012}%
  \BibitemOpen
  \bibfield  {author} {\bibinfo {author} {\bibfnamefont {J.}~\bibnamefont
  {Catani}}, \bibinfo {author} {\bibfnamefont {G.}~\bibnamefont {Lamporesi}},
  \bibinfo {author} {\bibfnamefont {D.}~\bibnamefont {Naik}}, \bibinfo {author}
  {\bibfnamefont {M.}~\bibnamefont {Gring}}, \bibinfo {author} {\bibfnamefont
  {M.}~\bibnamefont {Inguscio}}, \bibinfo {author} {\bibfnamefont
  {F.}~\bibnamefont {Minardi}}, \bibinfo {author} {\bibfnamefont
  {A.}~\bibnamefont {Kantian}},\ and\ \bibinfo {author} {\bibfnamefont
  {T.}~\bibnamefont {Giamarchi}},\ }\href
  {https://doi.org/10.1103/PhysRevA.85.023623} {\bibfield  {journal} {\bibinfo
  {journal} {Phys. Rev. A}\ }\textbf {\bibinfo {volume} {85}},\ \bibinfo
  {pages} {023623} (\bibinfo {year} {2012})}\BibitemShut {NoStop}%
\bibitem [{\citenamefont {Recati}\ \emph {et~al.}(2005)\citenamefont {Recati},
  \citenamefont {Fuchs}, \citenamefont {Peca},\ and\ \citenamefont
  {Zwerger}}]{Recati2005cfb}%
  \BibitemOpen
  \bibfield  {author} {\bibinfo {author} {\bibfnamefont {A.}~\bibnamefont
  {Recati}}, \bibinfo {author} {\bibfnamefont {J.~N.}\ \bibnamefont {Fuchs}},
  \bibinfo {author} {\bibfnamefont {C.~S.}\ \bibnamefont {Peca}},\ and\
  \bibinfo {author} {\bibfnamefont {W.}~\bibnamefont {Zwerger}},\ }\href
  {https://doi.org/10.1103/PhysRevA.72.023616} {\bibfield  {journal} {\bibinfo
  {journal} {Phys. Rev. A}\ }\textbf {\bibinfo {volume} {72}},\ \bibinfo
  {pages} {023616} (\bibinfo {year} {2005})}\BibitemShut {NoStop}%
\bibitem [{\citenamefont {Bordag}\ \emph {et~al.}(2009)\citenamefont {Bordag},
  \citenamefont {Fialkovsky}, \citenamefont {Gitman},\ and\ \citenamefont
  {Vassilevich}}]{Bordag2009}%
  \BibitemOpen
  \bibfield  {author} {\bibinfo {author} {\bibfnamefont {M.}~\bibnamefont
  {Bordag}}, \bibinfo {author} {\bibfnamefont {I.~V.}\ \bibnamefont
  {Fialkovsky}}, \bibinfo {author} {\bibfnamefont {D.~M.}\ \bibnamefont
  {Gitman}},\ and\ \bibinfo {author} {\bibfnamefont {D.~V.}\ \bibnamefont
  {Vassilevich}},\ }\href {https://doi.org/10.1103/PhysRevB.80.245406}
  {\bibfield  {journal} {\bibinfo  {journal} {Phys. Rev. B}\ }\textbf {\bibinfo
  {volume} {80}},\ \bibinfo {pages} {245406} (\bibinfo {year} {2009})},\
  \Eprint {https://arxiv.org/abs/0907.3242} {arXiv:0907.3242} \BibitemShut
  {NoStop}%
\bibitem [{\citenamefont {Rahi}\ \emph {et~al.}(2009)\citenamefont {Rahi},
  \citenamefont {Emig}, \citenamefont {Graham}, \citenamefont {Jaffe},\ and\
  \citenamefont {Kardar}}]{Rahi2009}%
  \BibitemOpen
  \bibfield  {author} {\bibinfo {author} {\bibfnamefont {S.~J.}\ \bibnamefont
  {Rahi}}, \bibinfo {author} {\bibfnamefont {T.}~\bibnamefont {Emig}}, \bibinfo
  {author} {\bibfnamefont {N.}~\bibnamefont {Graham}}, \bibinfo {author}
  {\bibfnamefont {R.~L.}\ \bibnamefont {Jaffe}},\ and\ \bibinfo {author}
  {\bibfnamefont {M.}~\bibnamefont {Kardar}},\ }\href
  {https://doi.org/10.1103/PhysRevD.80.085021} {\bibfield  {journal} {\bibinfo
  {journal} {Phys. Rev. D}\ }\textbf {\bibinfo {volume} {80}},\ \bibinfo
  {pages} {085021} (\bibinfo {year} {2009})},\ \Eprint
  {https://arxiv.org/abs/0908.2649} {arXiv:0908.2649} \BibitemShut {NoStop}%
\bibitem [{\citenamefont {Reichert}\ \emph {et~al.}(2019)\citenamefont
  {Reichert}, \citenamefont {Ristivojevic},\ and\ \citenamefont
  {Petkovi{\'{c}}}}]{Reichert2018}%
  \BibitemOpen
  \bibfield  {author} {\bibinfo {author} {\bibfnamefont {B.}~\bibnamefont
  {Reichert}}, \bibinfo {author} {\bibfnamefont {Z.}~\bibnamefont
  {Ristivojevic}},\ and\ \bibinfo {author} {\bibfnamefont {A.}~\bibnamefont
  {Petkovi{\'{c}}}},\ }\href@noop {} {\bibfield  {journal} {\bibinfo  {journal}
  {New J. Phys.}\ }\textbf {\bibinfo {volume} {21}},\ \bibinfo {pages} {053024}
  (\bibinfo {year} {2019})}\BibitemShut {NoStop}%
\bibitem [{\citenamefont {Dehkharghani}\ \emph {et~al.}(2018)\citenamefont
  {Dehkharghani}, \citenamefont {Volosniev},\ and\ \citenamefont
  {Zinner}}]{Dehkharghani2018}%
  \BibitemOpen
  \bibfield  {author} {\bibinfo {author} {\bibfnamefont {A.~S.}\ \bibnamefont
  {Dehkharghani}}, \bibinfo {author} {\bibfnamefont {A.~G.}\ \bibnamefont
  {Volosniev}},\ and\ \bibinfo {author} {\bibfnamefont {N.~T.}\ \bibnamefont
  {Zinner}},\ }\href {https://doi.org/10.1103/PhysRevLett.121.080405}
  {\bibfield  {journal} {\bibinfo  {journal} {Phys. Rev. Lett.}\ }\textbf
  {\bibinfo {volume} {121}},\ \bibinfo {pages} {80405} (\bibinfo {year}
  {2018})},\ \Eprint {https://arxiv.org/abs/1712.01538} {arXiv:1712.01538}
  \BibitemShut {NoStop}%
\bibitem [{\citenamefont {Schecter}\ and\ \citenamefont
  {Kamenev}(2014)}]{Schecter2014pmc}%
  \BibitemOpen
  \bibfield  {author} {\bibinfo {author} {\bibfnamefont {M.}~\bibnamefont
  {Schecter}}\ and\ \bibinfo {author} {\bibfnamefont {A.}~\bibnamefont
  {Kamenev}},\ }\href {https://doi.org/10.1103/PhysRevLett.112.155301}
  {\bibfield  {journal} {\bibinfo  {journal} {Phys. Rev. Lett.}\ }\textbf
  {\bibinfo {volume} {112}},\ \bibinfo {pages} {155301} (\bibinfo {year}
  {2014})}\BibitemShut {NoStop}%
\bibitem [{\citenamefont {Pavlov}\ \emph {et~al.}(2019)\citenamefont {Pavlov},
  \citenamefont {van~den Brink},\ and\ \citenamefont {Efremov}}]{Pavlov2019}%
  \BibitemOpen
  \bibfield  {author} {\bibinfo {author} {\bibfnamefont {A.~I.}\ \bibnamefont
  {Pavlov}}, \bibinfo {author} {\bibfnamefont {J.}~\bibnamefont {van~den
  Brink}},\ and\ \bibinfo {author} {\bibfnamefont {D.~V.}\ \bibnamefont
  {Efremov}},\ }\href {https://doi.org/10.1103/physrevb.100.014205} {\bibfield
  {journal} {\bibinfo  {journal} {Phys. Rev. B}\ }\textbf {\bibinfo {volume}
  {100}},\ \bibinfo {pages} {14205} (\bibinfo {year} {2019})}\BibitemShut
  {NoStop}%
\bibitem [{\citenamefont {Pavlov}\ \emph {et~al.}(2018)\citenamefont {Pavlov},
  \citenamefont {van~den Brink},\ and\ \citenamefont
  {Efremov}}]{Pavlov2018pmc}%
  \BibitemOpen
  \bibfield  {author} {\bibinfo {author} {\bibfnamefont {A.~I.}\ \bibnamefont
  {Pavlov}}, \bibinfo {author} {\bibfnamefont {J.}~\bibnamefont {van~den
  Brink}},\ and\ \bibinfo {author} {\bibfnamefont {D.~V.}\ \bibnamefont
  {Efremov}},\ }\href {https://doi.org/10.1103/PhysRevB.98.161410} {\bibfield
  {journal} {\bibinfo  {journal} {Phys. Rev. B}\ }\textbf {\bibinfo {volume}
  {98}},\ \bibinfo {pages} {161410(R)} (\bibinfo {year} {2018})}\BibitemShut
  {NoStop}%
\bibitem [{\citenamefont {Rodin}(2019)}]{Rodin2019}%
  \BibitemOpen
  \bibfield  {author} {\bibinfo {author} {\bibfnamefont {A.}~\bibnamefont
  {Rodin}},\ }\href {https://doi.org/10.1103/PhysRevB.100.195403} {\bibfield
  {journal} {\bibinfo  {journal} {Phys. Rev. B}\ }\textbf {\bibinfo {volume}
  {100}},\ \bibinfo {pages} {195403} (\bibinfo {year} {2019})}\BibitemShut
  {NoStop}%
\bibitem [{\citenamefont {Bruus}\ and\ \citenamefont
  {Flensberg}(2004)}]{Bruus2002}%
  \BibitemOpen
  \bibfield  {author} {\bibinfo {author} {\bibfnamefont {H.}~\bibnamefont
  {Bruus}}\ and\ \bibinfo {author} {\bibfnamefont {K.}~\bibnamefont
  {Flensberg}},\ }\href {https://doi.org/10.1088/0305-4470/38/8/B01} {\emph
  {\bibinfo {title} {Oxford Graduate Texts}}},\ \bibinfo {edition} {1st}\ ed.\
  (\bibinfo  {publisher} {Oxford University Press},\ \bibinfo {year}
  {2004})\BibitemShut {NoStop}%
\bibitem [{\citenamefont {Bezanson}\ \emph {et~al.}(2017)\citenamefont
  {Bezanson}, \citenamefont {Edelman}, \citenamefont {Karpinski},\ and\
  \citenamefont {Shah}}]{Bezanson2017}%
  \BibitemOpen
  \bibfield  {author} {\bibinfo {author} {\bibfnamefont {J.}~\bibnamefont
  {Bezanson}}, \bibinfo {author} {\bibfnamefont {A.}~\bibnamefont {Edelman}},
  \bibinfo {author} {\bibfnamefont {S.}~\bibnamefont {Karpinski}},\ and\
  \bibinfo {author} {\bibfnamefont {V.~B.}\ \bibnamefont {Shah}},\ }\bibfield
  {journal} {\bibinfo  {journal} {Society for Industrial and Applied
  Mathematics}\ }\textbf {\bibinfo {volume} {59}},\ \href
  {https://doi.org/10.1137/141000671} {10.1137/141000671} (\bibinfo {year}
  {2017})\BibitemShut {NoStop}%
\end{thebibliography}%

\end{document}